\documentclass[fleqn,usenatbib]{mnras}

\usepackage{newtxtext,newtxmath}
\usepackage{amsmath}

\usepackage[T1]{fontenc}

\DeclareRobustCommand{\VAN}[3]{#2}
\let\VANthebibliography\thebibliography
\def\thebibliography{\DeclareRobustCommand{\VAN}[3]{##3}\VANthebibliography}

\usepackage{graphicx}	
\usepackage{amsmath}	
\usepackage{setspace}
\newcommand{\formaldehyde}{$\mathrm{H_2CO}$}

\newcommand{\methanol}{$\mathrm{CH_3OH}$}
\newcommand{\htwo}{$\mathrm{H_2}$}
\newcommand{\gstate}{$1_{10}-1_{11}$}

\newcommand{\fstate}{$2_{11}-2_{12}$}
\newcommand{\threestate}{$3_{12}-3_{13}$}
\newcommand{\fourstate}{$4_{13}-4_{14}$}
\newcommand{\fivestate}{$5_{14}-5_{15}$}
\newcommand{\sixstate}{$6_{15}-6_{16}$}

\newcommand{\zerostate}{$1_{11}$}
\newcommand{\zerostateb}{$1_{10}$}

\newcommand{\firststateb}{$2_{11}$}

\newcommand{\taug}{$\tau_{4.8}$}
\newcommand{\tauf}{$\tau_{14.5}$}

\newcommand{\nhtwo}{$n_{\mathrm{H_2}}$}
\newcommand{\tk}{$T_k$}

\newcommand{\te}{$T_e$}

\newcommand{\whii}{$W_{\ion{H}{II}}$}
\newcommand{\rhii}{$R_{\ion{H}{II}}$}
\newcommand{\hii}{\ion{H}{II}}
\newcommand{\xform}{$X_{\mathrm{H_2CO}}$}

\title[]{Revisiting the formaldehyde masers II: effects of an \ion{H}{ii} region and beaming}

\author[D.J. van der Walt]{
D.J. van der Walt \thanks{E-mail: johan.vanderwalt@nwu.ac.za}
\\
Centre for Space Research, North-West University, Hoffman Street, 2520, Potchefstroom, South Africa
}

\date{Accepted 2024 September 16. Received 2024 September 13; in original form 2024 June 19}

\pubyear{2015}

\begin{document}
\label{firstpage}
\pagerange{\pageref{firstpage}--\pageref{lastpage}}
\maketitle

\begin{abstract}
  We present new results of a numerical study of the pumping of 4.8 GHz and 14.5 GHz maser
  of o-\formaldehyde{} in the presence of a free-free radiation field. It is shown that in
  the presence of a free-free radiation field inversion of not only the \gstate{}, but
  also the \fstate{} and other doublet state transitions occur. Further results are
  presented to illustrate how, as a consequence of the pumping scheme, the inversion of
  the \gstate{} and \fstate{} transitions respond to the free-free radiation fields
  associated with \ion{H}{II} regions with different emission measures and levels of
  geometric dilution with respect to the masing region. We also discuss the criticism
  raised in the past by various authors against the pumping of the 4.8 GHz \formaldehyde{}
  masers by a free-free radiation field. It is argued that the rarity of the
  \formaldehyde{} masers is not to be ascribed to the pumping scheme but to other factors
  such as, e.g., the evolution of the associated \ion{H}{II} region or the chemical
  evolution of the star forming region which determines the \formaldehyde{} abundance or a
  combination of both.
\end{abstract}




\section{Introduction}
\label{section:introduction}
Following the discovery of the first 4.8 GHz o-\formaldehyde{} (henceforth
\formaldehyde{}) maser in NGC 7538 by \citet{Downes1974}, \citet{Boland1981} presented the
first model to explain such maser emission. \citet{Boland1981} showed that the \gstate{}
transition can be inverted by the free-free radiation field of a compact \ion{H}{ii}
region. Since the model by \citet{Boland1981} was the only one for many years, various
authors tested the model against their observations. Authors such as, for example,
\citet{Gardner1986}, \citet{Mehringer1994}, \citet{Martin-Pintado1999},
\citet{Hoffman2003} and \citet{Araya2007c} concluded that the \formaldehyde{} masers known
at the time these papers were published, did not meet the conditions required by the model
of \citet{Boland1981} and, therefore, that the pumping of the 4.8 GHz \formaldehyde{}
masers by a free-free radiation field is not a viable mechanism to explain the
masers. Some of the reasons for dismissing the model of \citet{Boland1981} are, that the
emission measures (EM) of the \hii{} regions with which some of the 4.8 GHz
\formaldehyde{} masers are associated, are smaller than the lower limit of
$10^8\,\mathrm{pc\,cm^{-6}}$ for which \citet{Boland1981} found an inversion, and that
some of the masers are, according to the results of \citet{Boland1981}, offset too far
from the \hii{} regions to explain the inversion by a free-free radiation field. The
apparent failure of the \citet{Boland1981} model to explain the \formaldehyde{} masers
prompted \citet{Hoffman2007} to postulate that some rare collisional process may be
responsible for the pumping of the masers. Based on their observations of G339.88-1.26,
\citet{Chen2017a} argued in favour of the collisional excitation in a jet/outflow driven
shock of the \gstate{} \formaldehyde{} masers. Still more recently, \citet{McCarthy2022}
also express the opinion that the pumping mechanism for the population inversion of the
\gstate{} transition of \formaldehyde{} is poorly understood.

Another outstanding question to be addressed is the apparent absence of the 14.5 GHz
(\fstate{}) \formaldehyde{} maser which, according to the model of \citet{Boland1981},
should be associated with the 4.8 GHz maser if the pumping is through a free-free
radiation field. \citet{Hoffman2003} concluded that the non-detection or absence of the
14.5 GHz maser possibly indicates particular physical conditions for the excitation of the
4.8 GHz masers. Recently, \citet{Chen2017b} reported the detection of 14.5 GHz emission
from three high mass star forming regions which have associated 4.8 GHz \formaldehyde{}
masers. These authors consider the 14.5 GHz emission associated with NGC 7538 to be maser
emission, although \citet{Shuvo2021} argues that it might be thermal rather than maser
emission. Independent of whether the 14.5 GHz emission in NGC7538 is maser or thermal
emission, it is clear that, apart from understanding the inversion of the \gstate{}
transition, it is also necessary to understand the behaviour of the \fstate{} transition.

A further question related to the \formaldehyde{} \gstate{} masers is the rarity of these
masers. Counting individual masers, e.g. two components associated with NGC 7538, two with
G29.96-0.02, nine with Sgr B2, to date only 21 such masers have been detected in the
Galaxy \citep{Downes1974, Whiteoak1983, Mehringer1994, Pratap1994, Araya2005, Araya2006,
  Araya2008, Araya2015, Ginsburg2015, Chen2017a,McCarthy2022}. While the rarity of
  the masers may be related to the pumping mechanism, as proposed by \citet{Hoffman2007},
  there are also other possibilities which can explain the rarity of the masers. As one
of a number of possibilities, \citet{Ginsburg2015} also consider a short life-time of the
masers as a possible explanation of the rarity of the \gstate{} masers. Factors that might
determine the life-time of the masers are, for example, the chemical, dynamical and
radiative (dust and free-free) evolution of the star forming environment. The pumping
scheme, however, is determined exclusively by the molecular structure and not any external
macroscopic factors. Whether or not a certain transition is inverted depends on its
interaction (collisional and radiative) with its environment.

In a recent paper, \citet{vanderwalt2022} presented the results of a non-LTE study on the
inversion of the \gstate{} transition of \formaldehyde{}. The focus of these authors was
primarily on understanding the claim by \citet{baan2017} that the extragalactic megamasers
associated with three ultraluminous infrared (starburst) galaxies are radiatively pumped
by dust emission, as well as expanding on the results of \citet{vanderwalt2014} by
increasing the parameter space in kinetic temperature and \htwo{} density. These authors
showed that inversion of the \gstate{} transition can be achieved from collisions alone,
and that, in the presence of a far-infrared dust radiation field, inversion does not take
place when collisions are excluded. \citet{vanderwalt2022} also proposed a pumping scheme
which can explain why, in the case where the external radiation field is only that of
dust, collisions are required for the inversion of the \gstate{} transition.

The context of the present calculations is a pumping calculation with the aim to, first,
establish whether the pumping scheme proposed by \citet{vanderwalt2022} applies to the
case of excitation by a free-free radiation field. It is further investigated to what
extent the \gstate{} and \fstate{} transitions are inverted by different free-free
radiation fields and what the effect of beaming is. We also comment on past critique
against the viability of the pumping of the masers by a free-free radiation
field. Although it is necessary to try to reproduce the optical depths of the \gstate{}
transition as derived from observations \citep[for example in][]{Hoffman2003}, the aim of
the present calculations is not to explain individual sources.

\section{Numerical method}

The numerical method used for the present calculations is the same as that described in
detail by \citet{vanderwalt2021} and \citet{vanderwalt2022} and will therefore not be
repeated here. The molecular data used is also the same as that described in
\citet{vanderwalt2022}. We assumed a constant density and temperature \hii{} region to
calculate the free-free continuum radiation field given by
\begin{equation}
  F_\nu(T_e) = B_\nu(T_e)(1 - e^{-\tau_{\nu}}).
  \end{equation}
$B_\nu(T_e)$ is the Planck function, $T_e$ the electron temperature, and $\tau_\nu$ the
frequency dependent optical depth given by
\begin{equation}
    \tau_\nu =
    8.235\times10^{-2}\left(\frac{T_e}{\mathrm{K}}\right)^{-1.35}\left(\frac{\nu}{\mathrm{GHz}}
    \right)^{-2.1}\left(\frac{\mathrm{EM}}{\mathrm{pc\,cm^{-6}}}\right)
    \end{equation}
\citep{Wilson2009}. EM is the emission measure and is given by
\begin{equation}
  \mathrm{EM} = \int_{r_{\mathrm{inner}}}^{r_{\mathrm{outer}}}
  \left(\frac{n_e(r)}{\mathrm{cm^{-3}}}\right)^2d\left(\frac{r}{\mathrm{pc}}\right).
\end{equation}
with $n_e$ the electron density.

An important factor in the present calculations is the geometric dilution of the free-free
radiation field. The geometric dilution factor, $W_{\ion{H}{II}}$ is defined as the ratio
of the solid angle, $\omega$, subtended by the \hii{} region at the location of the masing
gas, relative to the total solid angle $4\pi$. Assuming a spherical \hii{} region and
expressing $\omega$ in terms of the radius, $R_{\hii{}}$, of the \hii{} region and the
distance, $r$, from the centre of the \hii{} region, the distance dependent dilution
factor is given by
\begin{figure}
  \begin{center}
    \includegraphics[width=\columnwidth]{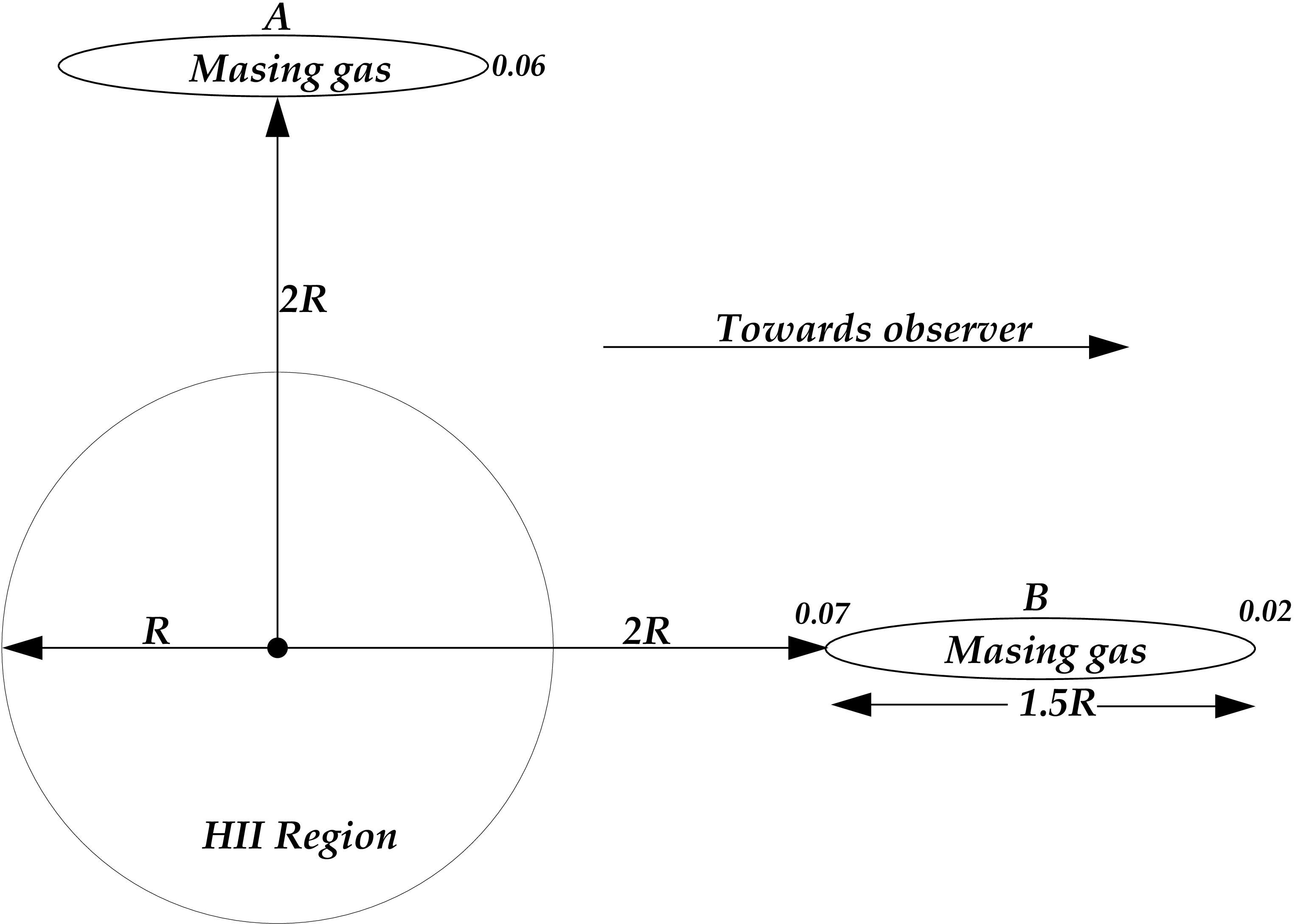}
    \caption
        {Schematic illustration of two possible positions of the masing gas relative to
          the \ion{H}{II} region with radius $R$. For case A the centre of the masing
          region, with linear dimension of 1.5R, is at a distance 2R from the centre of
          the \ion{H}{II} region which corresponds to \whii{} = 0.07. At the right hand
          end of the masing volume \whii{} = 0.06.  In case B the masing region is
          projected against the \ion{H}{II} region such that the left hand end is at
          distance 2R from the centre of the \ion{H}{II} region. The right hand end is at
          3.5R from the centre of the \ion{H}{II} region where \whii{} = 0.02.}
    \label{fig:geometry}
  \end{center}
\end{figure}

\begin{equation}
  W_{\ion{H}{II}}(r) = \frac{1}{2}\left\{1 - \sqrt{1 - \left(\frac{R_{\hii{}}}{r}\right)^2}\right\}.
  \label{eq:dilute1}
\end{equation}
For the case when $r \gg R_{\hii{}}$ it follows that
\begin{equation}
  W_{\ion{H}{II}}(r) \sim \frac{1}{4}\left(\frac{R_{\hii{}}}{r}\right)^2.
  \label{eq:dilute2}
\end{equation}
From Eq.\,\ref{eq:dilute1} it follows that $W_{\ion{H}{II}} = 0.5$ when $r = R_{\hii{}}$, and that $W_{\ion{H}{II}}$
is therefore restricted to values $ \leq 0.5$.  For all practical purposes one can say
that Eq.\,\ref{eq:dilute2} applies for $r > 5 R_{\hii{}}$ in which case $W_{\ion{H}{II}} < 0.01$.

 Within the framework of the escape probability method, and for an inverted
 transition, beaming is taken into account by modifying the escape probability
  \begin{equation}
    \beta_\nu = \frac{e^{\tau_\nu} - 1}{\tau_\nu}
  \end{equation}
  to become
  \begin{equation}
    \beta_\nu(\tau) = \frac{\Omega_\nu(\tau)}{4\pi}\frac{e^{\tau_\nu} - 1}{\tau_\nu}
    \label{eq:beaming}
    \end{equation}
  where $\Omega_\nu$ is the frequency dependent beaming angle \citep[see Section 5.3
    of][]{elitzur1992}. According to \citet[][Section 5.1.3]{elitzur1992}, the beaming
  solid angle varies as $\ell^{-1}$ for unsaturated masers and as $\ell^{-2}$ for
  saturated masers, where $\ell$ is the length along the ray path. Since $\tau \propto
  \ell$, it was assumed that
  \begin{equation}
    \frac{\Omega_\nu(\tau_\nu)}{4\pi} = \frac{1}{\left(\alpha_\nu \tau_\nu + 1 \right)^{2}}
    \label{eq:beamfact}
  \end{equation}
  as an approximation of such behaviour. Thus,
  \begin{equation}
    \beta_\nu(\tau) = \frac{1}{\left(\alpha_\nu \tau_\nu + 1 \right)^{2}}\frac{e^{\tau_\nu} - 1}{\tau_\nu}.
    \label{eq:finbeaming}
  \end{equation}
  It should be noted, however, that $e^{\tau}$ increases faster than what $\tau^{-2}$
  decreases for $\tau \gtrsim 1$ and, therefore, that $\beta_\nu(\tau)$ increases toward
  larger values of $\tau$ rather than to decrease. Inspection of the behaviour of
  $\beta_\nu(\tau)$ (as in Eq.\,\ref{eq:finbeaming}) has further shown that
  $\beta_\nu(\tau)$ has a power law behaviour of the form $a\tau^{-\gamma}$ for $0.4 <
  \tau \le 1$ for $\alpha \ge 1$. To avoid the escape probability to increase for $\tau >
  1$, it was assumed that the power law behaviour continues for $\tau > 1$. The power law
  index, $\gamma$, depends on $\alpha_\nu$; for example, $\gamma = -0.41, -1.15$, and
  $-1.36$ respectively for $\alpha = 1,~5$ and 10.
  
 Although it is not the explicit intention of the calculations to model specific
    maser sources, it is necessary to use beaming angles that correspond to observed 4.8
    GHz maser spot sizes. Consider therefore a maser spot at a distance $d$ from the
    observer and which has an angular size of $\theta$ radians. The surface area of the
    spot is then $A = \pi d^2\theta^2 /4$. On the other hand, for a maser path length
    $\ell_m$ and a beaming angle $\Omega$ given by Eq.\,\ref{eq:beamfact}, the cross
    sectional area of the cone is $A = 4\pi\ell_m^2/(\alpha\tau + 1)^2$. Equating the two
    areas, it follows that
  \begin{equation}
    \alpha\tau = \frac{4\ell_m}{\theta d} - 1
    \label{eq:alphatau}
  \end{equation}
  To illustrate the application of Eq.\,\ref{eq:alphatau} consider, for example, the case
  of G29.96-0.02. \citet{Hoffman2003} measured the angular size of the two maser
  components to be $\sim$19 milli-arcseconds at a distance of 6.5 kpc. For a maser path
  length of $7\times 10^{16}$ cm (to be discussed later) we have $\alpha\tau =
  150$. Choosing, for example, $\alpha = 15$ implies that for $\tau = 10$ the angular
  diameter of the spot will agree with the observed angular diameter.
  
 The present calculations are done within the framework of the escape probability method
 which requires that the physical conditions, e.g. density, temperature and the free-free
 radiation field, be uniform over the length of the masing column. To meet this
 requirement all parts of the masing volume of gas should be almost equidistant from the
 \ion{H}{II} region. The implied geometry is therefore a volume of gas with length
 $\ell_m$ illuminated from the side with the free-free emission from a hyper- or
 ultra-compact \ion{H}{II} region, schematically illustrated as case A in
 Fig.\,\ref{fig:geometry}. The masing region can also be located as in case B. In this
 scenario there is a gradient in the geometric dilution of the free-free radiation field
 across the length of the masing region which means non-uniform pumping rates between the
 two end points of the masing region. Case B is certainly the more realistic scenario but
 cannot be dealt with using the escape probability method. Case A is, however, sufficient
 to investigate to what extent the free-free radiation field can lead to the inversion of
 the \gstate{} and \fstate{} transitions. Case A is qualitatively similar to the LVG
 method for a uniform source as was used by \citet{Cragg2002} to study the pumping of
 \methanol{} masers.

\section{Pumping scheme}

In brief, the pumping scheme proposed by \citet{vanderwalt2022} is as follows: In the
presence of a far-infrared radiation field radiative excitation out of \zerostate{} into
the ladders of lower and upper doublet states is asymmetric. This follows from the fact
that although the Einstein B-coefficients for the transitions $1_{11} \rightarrow 1_{10}$
(4.8 GHz) and $1_{11} \rightarrow 2_{12}$ (140.8 GHz) are almost equal, the spectral
energy density at 140.8 GHz is significantly larger than at 4.8 GHz for a typical
far-infrared radiation field of the form $F_\nu(T_d)= (1 -
\exp(-(\nu/\nu_0)^p))B_\nu(T_d)$ with e.g. $T_d \sim$ 50 to 100 K, $p = 1.8$, and $\nu_0
\sim 10^{12}$ Hz. Thus, radiative excitation out of \zerostate{} will therefore populate
the ladder of lower double states faster than the ladder of upper doublet states. Using
the collision coefficients of the Leiden Atomic and Molecular Database \citep{Schoier2005}
it is rather straight forward to show, for collisions with both $\mathrm{o-H_2}$ and
$\mathrm{p-H_2}$, that there is a slight asymmetry which favours collisional excitation
from \zerostate{} into the ladder of lower doublet states and that the asymmetry increases
with increasing kinetic temperature. To create an inversion of the \gstate{} transition,
it is necessary to transfer molecules from the ladder of lower doublet states to the
ladder of upper doublet states. For the case where excitation is through a far-infrared
radiation field and collisions, the transfer of molecules from the ladder of lower doublet
states to the ladder of upper doublet states is collisional as was shown by
\citet{vanderwalt2022}.

\begin{figure}
  \begin{center}
    \includegraphics[width=\columnwidth]{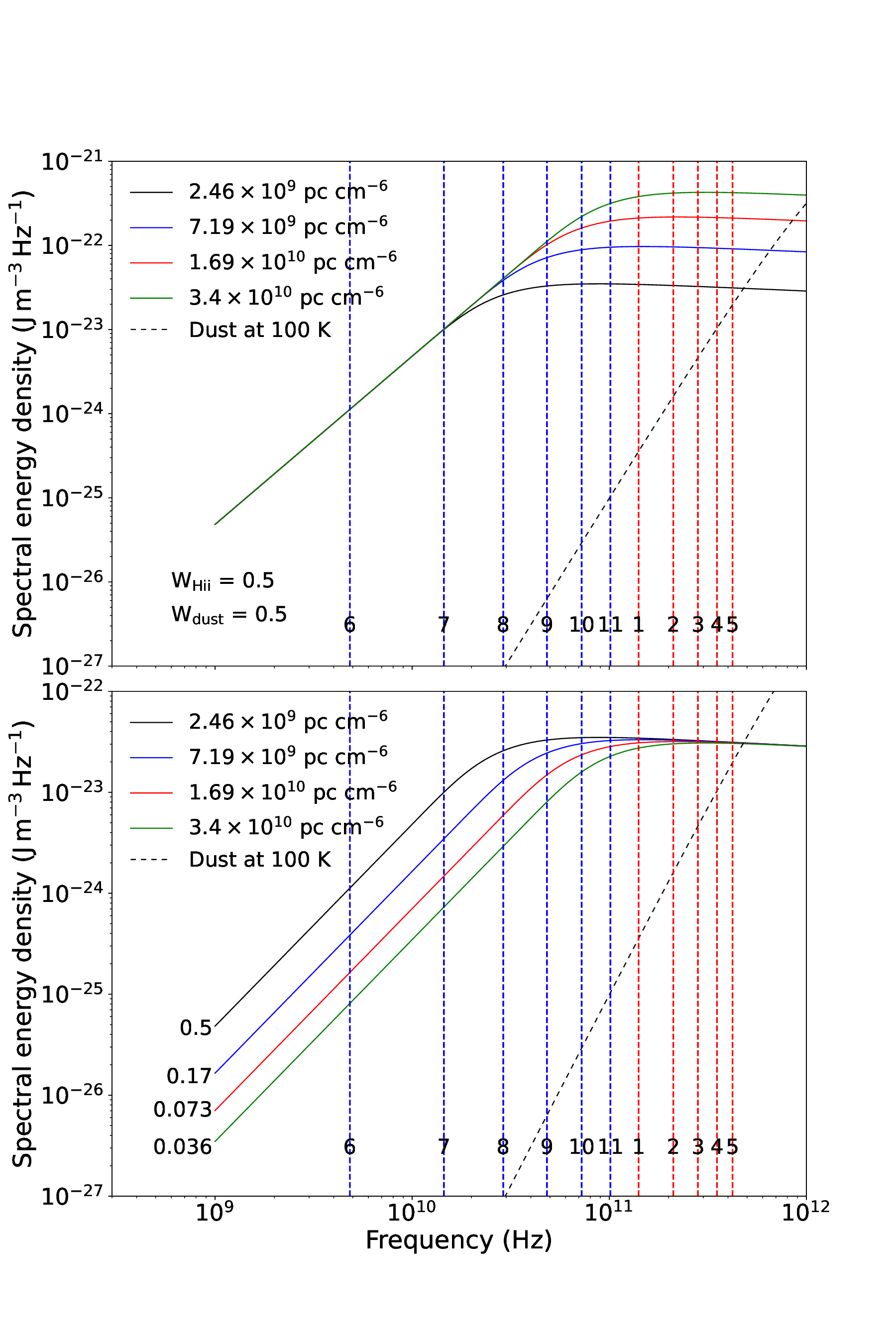}
    \caption{{\it Upper panel:} Theoretical spectral energy density distributions for four
      \ion{H}{II} regions with emission measures and dilution factor as shown. The
      vertical dashed red lines numbered 1 to 5 correspond respectively to the first five
      $\Delta J = 1, K_a = 1, \Delta K_c = 1$ transitions in the ladder of lower doublet
      states with frequencies 140.84, 211.21, 281.53, 351.77 and, 421.92 GHz. The vertical
      dashed blue lines numbered 6 to 11 correspond respectively to the first five $\Delta
      J = 0, K_a = 1, \Delta K_c = -1$ transitions from the ladder of lower doublet states
      to the ladder of upper doublet states with frequencies 4.83, 14.49, 28.97, 48.28
      and, 72.41 GHz. {\it Bottom panel:} Same as for the upper panel but with the
      spectral energy densities normalized at $10^3$ GHz to that for EM = $2.46 \times
      10^{9}\,\mathrm{pc\, cm^{-6}}$ with \whii{} = 0.5. The resulting dilution factors
      associated with each emission measure are shown on the graph.}
  \label{fig:emh2co}
  \end{center}
\end{figure}

To illustrate the pumping scheme in the presence of a free-free radiation field (but also
for later examples) four emission measures were chosen such that the turnover frequencies
of the free-free radiation fields correspond to the frequencies for the \threestate{}
(28.9 GHz), \fourstate{} (48.3 GHz), \fivestate{} (72.4 GHz) and, \sixstate{} (101.3 GHz)
doublet transitions. The turnover frequencies therefore correspond to transitions where
molecules can be transfered from the ladder of lower doublet states to the ladder of upper
doublet states. The associated emission measures (with units of $\mathrm{pc\,cm^{-6}}$)
are respectively $2.46 \times 10^9$, $7.19 \times 10^9$, $1.69 \times 10^{10}$ and $3.41
\times 10^{10}$. The respective spectral energy density distributions are shown in
Fig.\,\ref{fig:emh2co} where, in the upper panel \whii{} = 0.5 for all four emission
measures and in the lower panel for different dilution factors such that the spectral
energy densities at 1000 GHz are the same as for EM = $2.46 \times
10^9\,\mathrm{pc\,cm^{-6}}$ with \whii{} = 0.5. This means that \whii{} = 0.171 for EM =
$7.19 \times 10^9\, \mathrm{pc\, cm^{-6}}$, \whii{} = 0.073 for EM = $1.69 \times
10^{10}\, \mathrm{pc\, cm^{-6}}$ and, \whii{} = 0.036 for EM = $3.41 \times 10^{10}\,
\mathrm{pc\, cm^{-6}}$. The upper and lower panels of Fig.\,\ref{fig:emh2co} therefore
represent two special cases where, for the upper panel, the excitation rates (which are
directly proportional to the spectral energy density, Eq.\,\ref{eq:transitionrates}) for
$\Delta J = 1, \Delta K_c = 1$ transitions within the ladder of lower doublet states
(lines 1 to 5) range approximately over an order of magnitude from EM = $2.46 \times
10^{9}\,\mathrm{pc\,cm^{-6}}$ to EM = $3.41 \times 10^{10}\,\mathrm{pc\,cm^{-6}}$. On the
other hand, the $\Delta J = 0, \Delta K_c = -1$ transition rates for the three lowest
doublet states (lines 6, 7, 8) are approximately the same for the four emission
measures. In the lower panel the situation is reversed, that is, the excitation rates
within the ladder of lower doublet states (lines 1 to 5) are practically the same for the
four emission measures while there are marked differences for the $\Delta J = 0, \Delta
K_c = -1$ transitions for the transfer of molecules from the ladder of lower doublet
states to the ladder of upper doublet states. Also shown in both panels is the spectral
energy density distribution for a far-infrared radiation field for $T_d$ = 100 K and a
dilution factor of 0.5.

For radiative excitation out of \zerostate{} there are, as mentioned above, two
possibilities, viz. $1_{11} \rightarrow 1_{10}$ (4.8 GHz - dashed blue vertical line 6)
and $1_{11} \rightarrow 2_{12}$ (140.8 GHz - dashed red vertical line 1). Due to the
significant difference in the spectral energy densities at these two frequencies,
radiative excitation rate from $1_{11} \rightarrow 2_{12}$ is higher than from $1_{11}
\rightarrow 1_{10}$, resulting in a faster population of the ladder of lower doublet
states compared to the ladder of upper doublet states. Further excitation within the
ladder of lower doublet states, for example, $2_{12} \rightarrow 3_{13} \rightarrow 4_{14}
\rightarrow 5_{15} \rightarrow 6_{16}$ are facilitated through absorption of photons
respectively at frequencies indicated by the vertical dashed red lines 2, 3, 4, 5. The
transfer of molecules from the ladder of lower doublet states to the ladder of upper
doublet states, for example $2_{12} \rightarrow 2_{11}, 3_{13}\rightarrow 3_{12}$ etc, is
through the absorption of photons at frequencies indicated by the vertical dashed blue
lines 7 to 11.

The important point to be established is whether the transfer of molecules from the ladder
of lower doublet states to the ladder of upper doublet states is essential for an
inversion of the \gstate{} transition. The radiative transition rate between a lower and
an upper state is given by
\begin{equation}
  \Gamma_{lu}=B_{lu}U_{ul} = \frac{g_u}{g_l}\frac{c^3}{8\pi h \nu^3_{ul}}A_{ul}U_{ul}.
  \label{eq:transitionrates}
\end{equation}
where $U_{ul}$ is the spectral energy density of the free-free radiation field at the
frequency $\nu_{ul}$.  $\Gamma_{lu}$ can be changed by changing either $A_{ul}$ or
$U_{ul}$. To test whether the inversion of the \gstate{} transition depends on the rate at
which molecules are transfered from the ladder of lower doublet states to the ladder of
upper doublet states ($\Delta J = 0, \Delta K_c = -1$), it is necessary to only change the
rates of these transitions while keeping the rates for all other transitions the same. The
simplest way to do this is by adjusting, with the exception of the $1_{11}\rightarrow
1_{10}$ transition (through which molecules are radiatively excited into the ladder of
upper doublet states), the Einstein A-coefficients for the $\Delta J = 0, \Delta K_c = \pm
1$ transitions rather than modifying the radiation field.

\begin{figure}
  \begin{center}
    \includegraphics[width=\columnwidth]{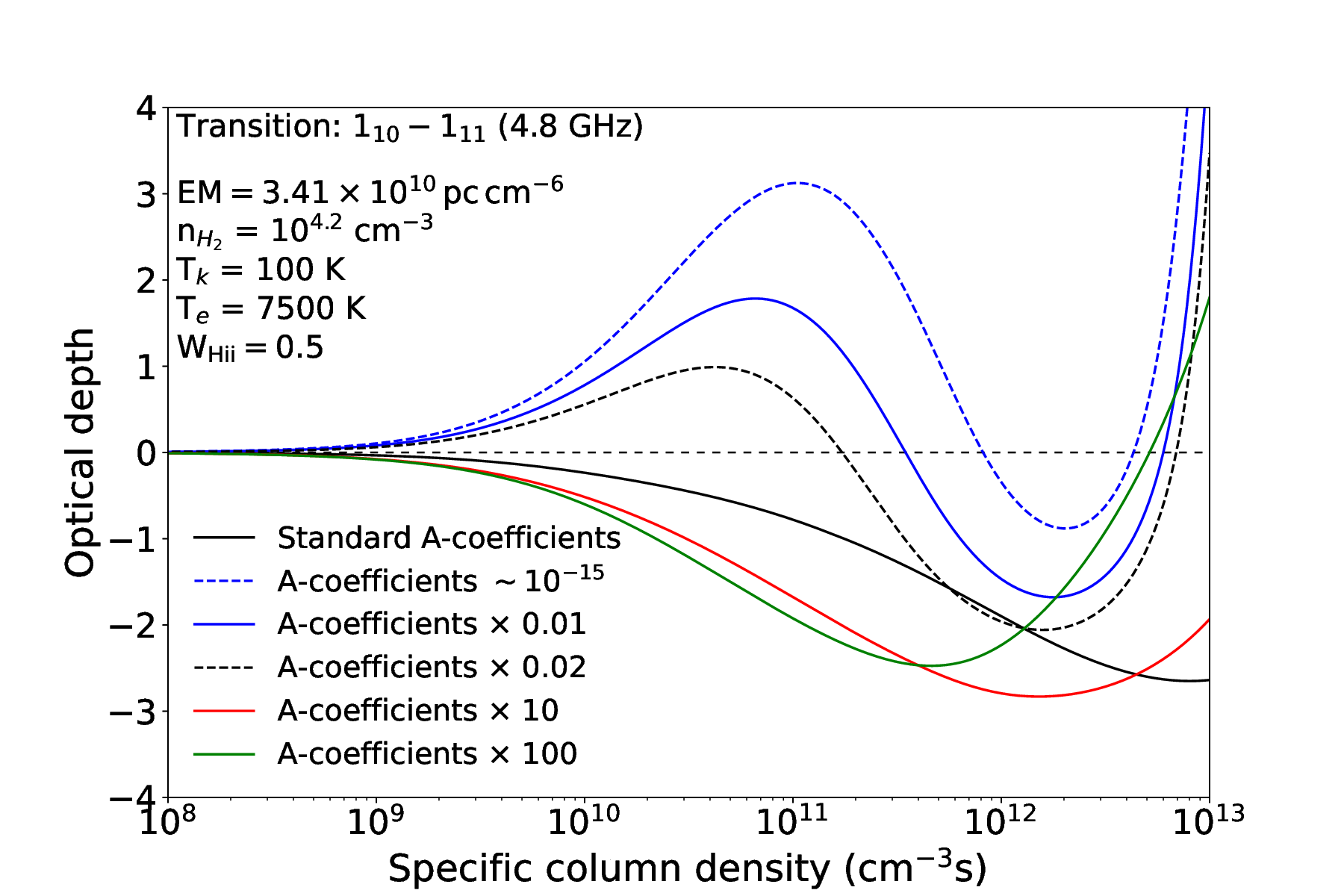}
  \caption {Comparison of the variation of the optical depth of the \gstate{} transition as function of specific
    column density where the Einstein A-coefficients for the doublet state transitions
    ($\Delta J = 1, \Delta K_c = \pm 1$) have been adjusted to illustrate the effect of
    changing the rate of transfer of molecules from the ladder of lower doublet to the
    ladder of upper doublet states.}
    \label{fig:acoefdep}
    \end{center}
\end{figure}

Figure\,\ref{fig:acoefdep} shows an example of the variation of \taug{} with specific
column density when (a) the ``standard'' A-coefficients for the $\Delta J = 0, \Delta K_c
= \pm 1$ transitions are used (black line), (b) when all the coefficients are set to be of
the order of $10^{-15}\,\mathrm{s^{-1}}$ (dashed blue line) (c) when the standard
coefficients are reduced by a factor of 100 (solid blue line), (d) when the standard
coefficients are reduced by a factor of 50 (dashed black line), (e) when standard
coefficients are increased by one and two orders of magnitude above their standard values
(red and green lines respectively). The A-coefficients for all the other allowed
transitions were kept at their ``standard'' values. Reducing the A-coefficients for the
$\Delta J = 0, \Delta K_c = \pm 1$ transitions to be of the order of
$10^{-15}\,\mathrm{s^{-1}}$, implying that there is virtually no exchange of molecules
between the ladder of lower doublet states and the ladder of upper doublet states, has as
consequence that there is no inversion of the \gstate{} transition for specific column
densities less than $\sim 8 \times 10^{11}\,\mathrm{cm^{-3}\,s}$. The small inversion that
follows can be shown to be due to collisional transfer of molecules between the ladders of
lower and upper doublet states.  The trends in the behaviour of \taug{} for the different
cases shown in Fig.\,\ref{fig:acoefdep} clearly show that the inversion of the \gstate{}
transition depends on the transfer of molecules from the ladder of lower doublet states to
the ladder of upper doublet states. The pumping scheme proposed by \citet{vanderwalt2022},
as explained above, therefore also applies when excitation is due to a free-free radiation
field, the difference being that in the presence of a free-free radiation field the
pumping is dominantly radiative.

Comparing the spectral energy density distributions in both panels of
Fig.\,\ref{fig:emh2co} clearly shows that for the far-infrared radiation field the
spectral energy densities responsible for the excitation from \zerostate{} into the ladder
of lower doublet states (dashed vertical red lines 1 to 5 ) are significantly
smaller (in some cases more than an order of magnitude) than for the free-free radiation
fields. It is also seen for the far-infrared radiation field, that the spectral energy
densities at the frequencies responsible for the transfer of molecules from the ladder of
lower doublet states to the ladder of upper doublet states (dashed vertical blue lines 6
to 11) are many orders of magnitude smaller than for the free-free radiation fields. The
dust radiation field therefore plays a completely insignificant role in the pumping of the
masers in regions where the free-free radiation field dominates.

\label{section:pumping}

\section{Illustrative examples}

\subsection{Basic results without beaming}

\begin{figure}
  \begin{center}
    \includegraphics[width=\columnwidth]{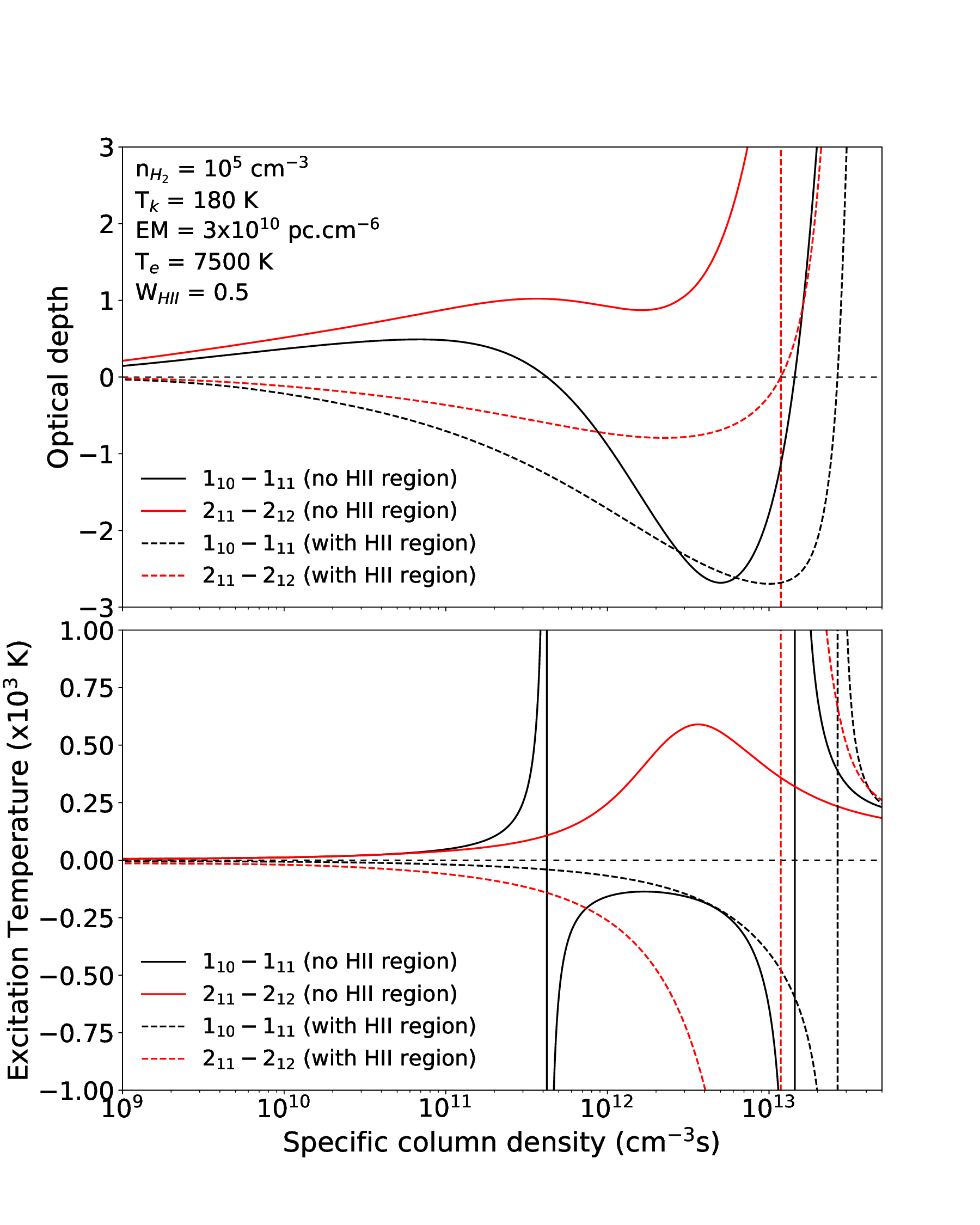}
  \caption{\textit{Upper panel:} Comparison of the variation of the optical depth as function of specific
    column density for the \gstate{} and \fstate{} transitions in the absence of a
    free-free radiation field with the case when a free-free radiation field is present.
      \textit{Bottom panel:} Comparison of the variation of the excitation temperature for
      the \gstate{} and \fstate{} transitions.} 
    \label{fig:tautex01}
    \end{center}
\end{figure}

The basic result of the calculational procedure to solve for the level populations is the
variation of the level populations, and therefore also quantities such as the optical
depth and the excitation temperature, as a function of the \formaldehyde{} specific column
density as described in \citet{vanderwalt2021}. As a first illustrative result, the
variation of the optical depth for the \gstate{} (\taug{}) and \fstate{} (\tauf{})
transitions in the presence and absence of a free-free radiation field are compared in the
upper panel of Fig.\,\ref{fig:tautex01}. The results were obtained for \nhtwo{} =
$10^{5}~\mathrm{cm^{-3}}$, \tk{} = 180 K, an \formaldehyde{} abundance (\xform{}) of $5.0
\times 10^{-6}$, a maser linewidth of $0.3~\mathrm{km\,s^{-1}}$, EM = $3 \times
10^{10}~\mathrm{pc\,cm^{-6}}$ and \whii{} = 0.5. The lower panel shows the corresponding
variation of the excitation temperatures for the two transitions where negative excitation
temperatures indicate a population inversion. It is seen that in the absence of a
free-free radiation field, the \gstate{} transition is not inverted up to a specific
column density of $\sim 4 \times 10^{11}\,\mathrm{cm^{-3}\,s}$. The \fstate{} transition
is not inverted, which was more generally found to hold for all combinations of \nhtwo{}
and \tk{} where the \gstate{} transition is inverted in the absence of a free-free
radiation field. In the presence of a strong free-free radiation field (minimum geometric
dilution) both transitions are seen to be inverted, even at those specific column
densities where the \gstate{} transition is not inverted in the absence of a free-free
radiation field.

\begin{figure}
  \begin{center}
    \includegraphics[width=\columnwidth]{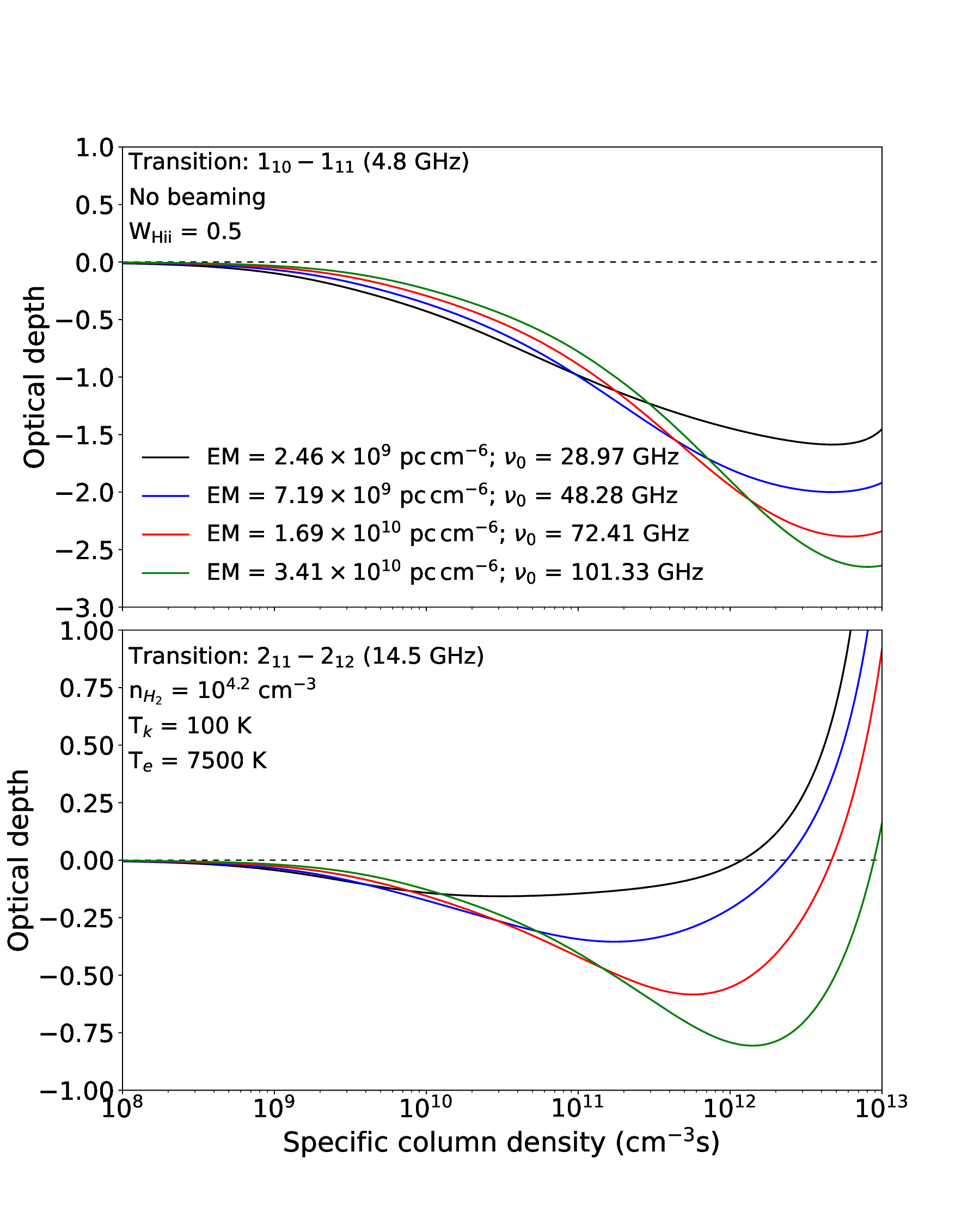}
  \caption{ Variation of the optical depths for the \gstate{} (upper panel) and \fstate{}
    (lower panel) transitions as a function specific column density for the four emission
    measures and a dilution factor of 0.5 as shown in the upper panel of Fig.\,\ref{fig:emh2co}.}
    \label{fig:emdep01}
    \end{center}
\end{figure}

\begin{figure}
  \begin{center}
    \includegraphics[width=\columnwidth]{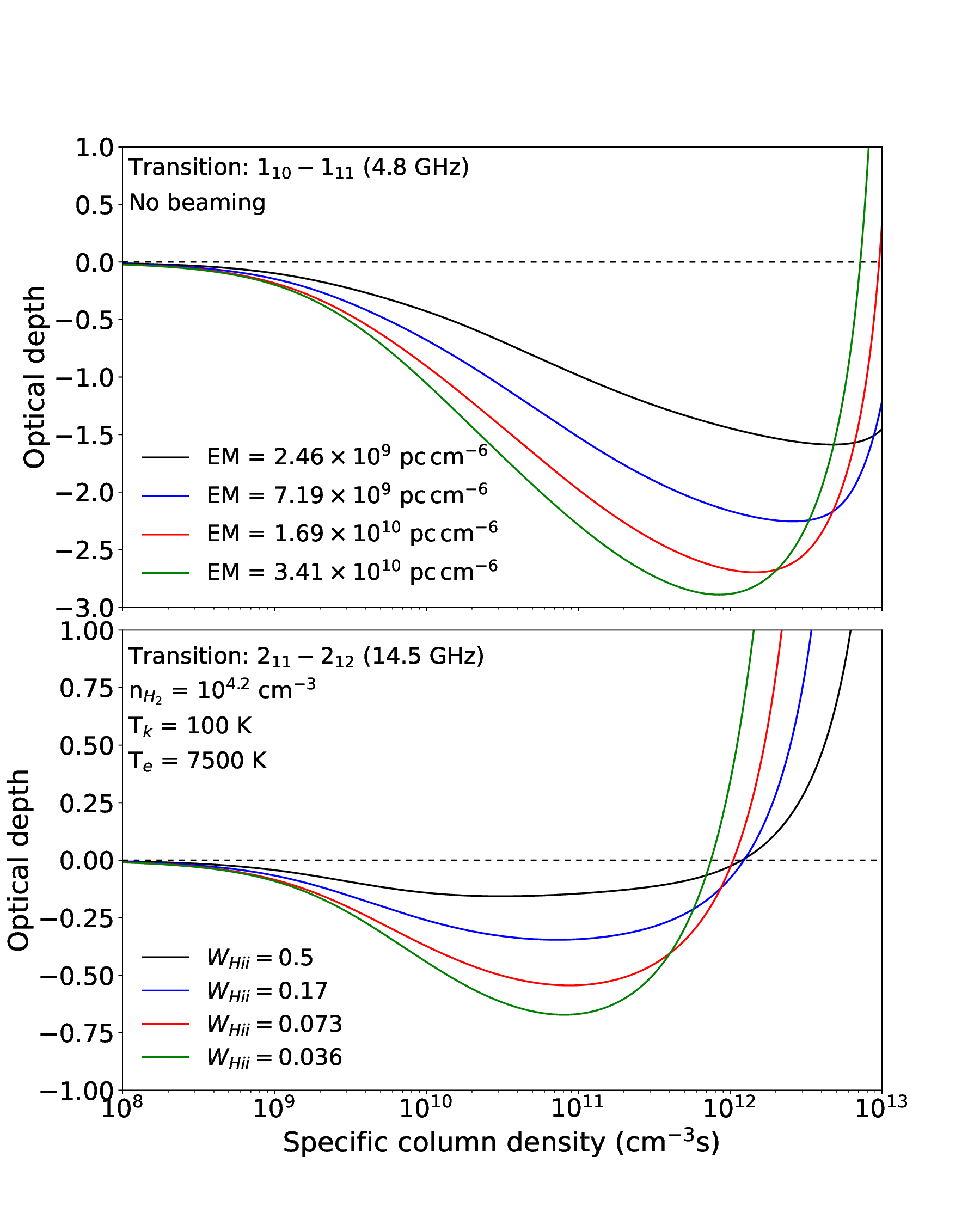}
    \caption{Variation of the optical depths for the \gstate{} (upper panel) and \fstate{}
      (lower panel) transitions as a function specific column density for the four
      emission measures and dilution factors as shown in the lower panel of
      Fig.\,\ref{fig:emh2co}. }
    \label{fig:emdep02}
    \end{center}
\end{figure}

Since, as was shown in Section\,\ref{section:pumping}, the inversion of the \gstate{}
transition can be achieved through the pumping by a free-free radiation field, it can be
expected that \taug{} and \tauf{} depend on the emission measure and the dilution
factor, \whii{}.  The variation of \taug{} and \tauf{} as a function of the specific
column density for the two scenarios shown in the upper and lower panels of
Fig.\,\ref{fig:emh2co} is shown respectively in Figs.\,\ref{fig:emdep01} and
\ref{fig:emdep02} with \nhtwo{} = $10^{4.2}\,\mathrm{cm^{-3}}$, \tk{} = 100 K and, \te{} =
7500 K. It is worth keeping in mind that, from Eq.\,\ref{eq:dilute1}, the dilution factors
of 0.5, 0.171, 0.073 and 0.036 correspond respectively to distances of 1.0\rhii{},
1.3\rhii{}, 1.9\rhii{} and, 2.7\rhii{} from the centre of the \hii{} region.


Comparison of Figs.\,\ref{fig:emdep01} and \ref{fig:emdep02} shows some obvious
differences. For example, in Fig.\,\ref{fig:emdep01} it is seen that there is very little
difference in \taug{} for the different emission measures for specific column densities up
to $\sim 3 \times 10^{11}\,\mathrm{cm^{-3}\,s}$.  This can partly be ascribed to the
transition rates being equal (or approximately equal) for the first three $\Delta J = 0,
\Delta K_c = -1$ transitions (lines 6, 7 and 8 in the upper panel of
Fig.\,\ref{fig:emh2co}) although the excitation rates into and within the ladder of lower
doublet states (lines 1 to 5 in the upper panel of Fig.\,\ref{fig:emh2co}) differ
significantly. On the other hand, when the dilution factors for the four emission measures
are different, as in the lower panel of Fig.\,\ref{fig:emh2co}, the variation of \taug{}
with specific column density as shown in Fig.\,\ref{fig:emdep02} is quite different from
that in Fig.\,\ref{fig:emdep01}.  In fact, it is seen in Fig.\,\ref{fig:emdep02} that the
maximum of \taug{} for EM = $3.41 \times 10^{10}\,\mathrm{pc\,cm^{-6}}$ at a distance of
$2.7R_{\ion{H}{II}}$ is greater than for EM = $2.46 \times 10^{9}\,\mathrm{pc\,cm^{-6}}$
with the masing region located at the edge of the \ion{H}{II} region. Also, closer
inspection of the upper panel of Figs.\,\ref{fig:emdep01} shows that \taug{} is greater
for EM = $2.46 \times 10^{9}\,\mathrm{pc\,cm^{-6}}$ than for EM = $3.41 \times
10^{10}\,\mathrm{pc\,cm^{-6}}$ for specific column densities less than $\sim\,3 \times
10^{11}\,\mathrm{cm^{-3}\,s}$. At first sight, this behaviour appears to be counter
intuitive since, if the masing region is closer to the \ion{H}{II} region, pumping might
be expected to be more effective to establish a larger inversion of the \gstate{}
transition.

\begin{figure}
  \begin{center}
    \includegraphics[width=\columnwidth]{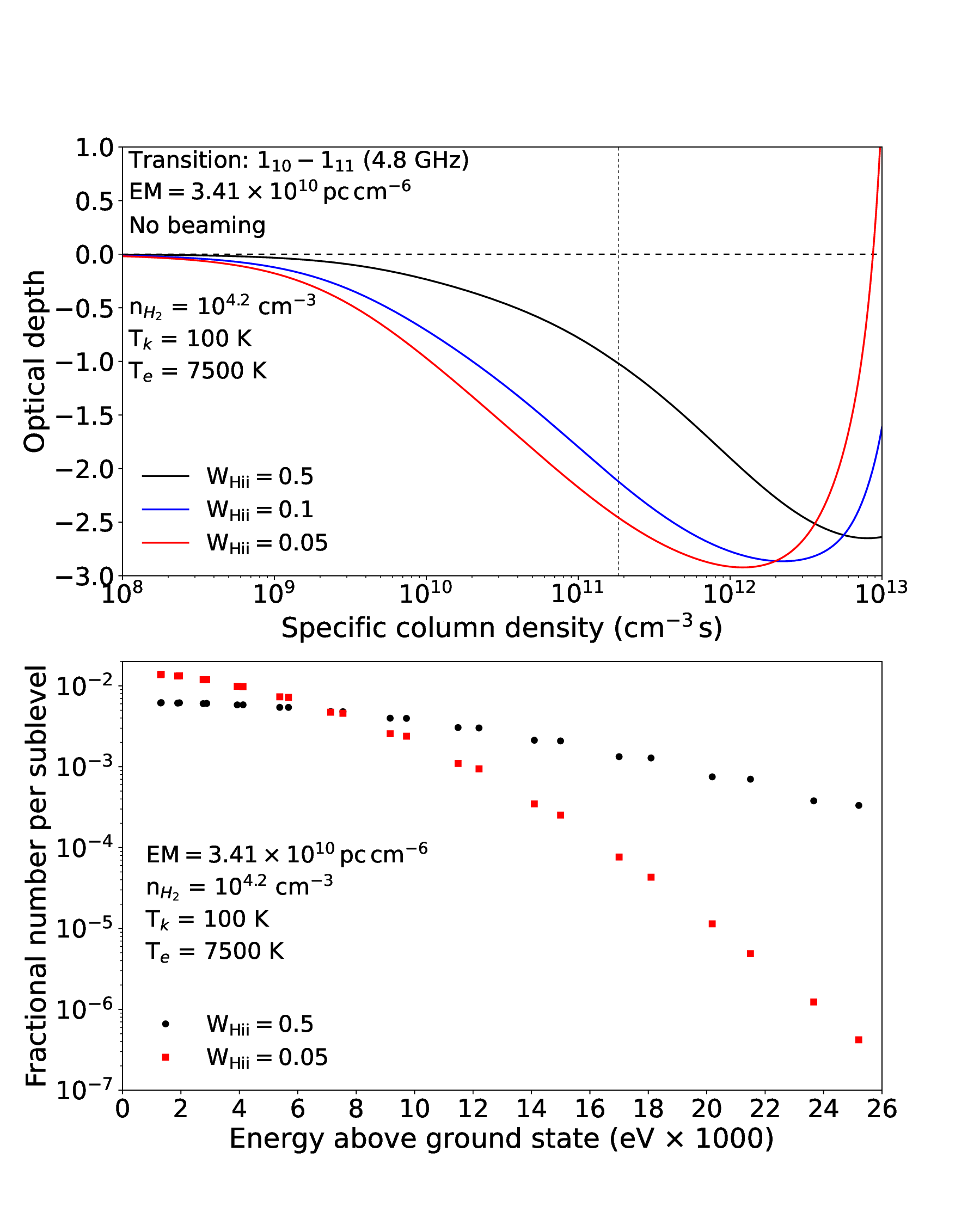}
    \caption{ \textit{Upper panel:} Variation of the optical depth of the \gstate{} transition
      to illustrate the behaviour of \taug{} for different dilution factors. The vertical
      dashed line is located at a specific column density of $1.8 \times
      10^{11}\,\mathrm{cm^3\,s^{-1}}$ which corresponds to a maser pathlength of $7 \times
      10^{16}$ cm for $\mathrm{n_{H_2} = 10^{4.2}\,\mathrm{cm^{-3}}}$, a \formaldehyde{}
      abundance of $5 \times 10^{-6}$ and, a velocity width of 300
      $\mathrm{m\,s^{-1}}$. \textit{Lower panel:} Solution for the level populations at a
      maser pathlength of $7 \times 10^{16}$ cm for the dilution factors as shown on the graph. 
    }
    \label{fig:whiidep}
    \end{center}
\end{figure}

Some insight into this behaviour can be gained by considering the radiative excitation
rates between states within the ladders of lower and upper doublet states ($\Delta J = 1,
\Delta K_c = \pm 1$) as well as between the lower to the upper doublet states ($\Delta J =
0, \Delta K_c = \mp 1$).  To illustrate this we show the variation of \taug{} on the
specific column density for \whii{} = 0.5, 0.1, and 0.05 for a free-free radiation field
with EM = $3.41 \times 10^{10}\,\mathrm{pc\,cm^{-6}}$ in the upper panel of
Fig.\,\ref{fig:whiidep}. The values of the other parameters are as shown in the
Figure. For a maser pathlength of $7 \times 10^{16}$ cm (at the specific column density
indicated by vertical dashed line in the upper panel; see Section\, \ref{section:beaming})
it is seen that $\lvert$\taug{}$\rvert_{\mathrm{W_{Hii}=0.5}} < $
$\lvert$\taug{}$\rvert_{\mathrm{W_{Hii}=0.1}}<$
$\lvert$\taug{}$\rvert_{\mathrm{W_{Hii}=0.05}}$. Calculation of the radiative excitation
rates for the $\Delta J = 1, \Delta K_c = 1$ transitions when \whii{} = 0.5, gives rates
between $\sim 0.019\,\mathrm{s^{-1}}$ and $\sim 0.023\,\mathrm{s^{-1}}$. This translates
to characteristic `lifetimes' for a \formaldehyde{} molecule in the lower states of these
transitions to be between 42 and 51 seconds.  The corresponding transition rates for a
molecule in the ladder of lower doublet states to make a transition to the associated
upper doublet state ($\Delta J = 0, \Delta K_c = -1$) fall between $6 \times
10^{-5}\,\mathrm{s^{-1}}$ and $8 \times 10^{-4}\,\mathrm{s^{-1}}$, which translates to
characteristic lifetimes between 1250 and 16700 seconds. The effect of the much higher
transition rates (shorter lifetimes) for $\Delta J = 1, \Delta K_c= 1$ transitions
compared to the $\Delta J = 0, \Delta K_c = -1$ transitions is, therefore, that molecules
excited into the ladder of lower doublet states will tend to stay in the ladder of lower
doublet states with only a small probability of making a transition to the ladder of upper
doublet states. Even if there is any population inversion for the \gstate{} transition,
the high radiative excitation rate for \zerostateb{}$\rightarrow$\firststateb{} will
further tend to reduce the magnitude of the inversion. On the other hand, for \whii{} =
0.05, the transition rates are an order of magnitude smaller with the characterisc
lifetimes for the $\Delta J = 1, \Delta K_c = 1$ transitions, ranging between 430 and 513
seconds which increases the probability for a molecule in the ladder of lower doublet
states to be transfered to the ladder of upper doublet states.  The resulting level
populations for the two dilution factors for a maser pathlength of $7 \times 10^{16}$ cm
are compared in the bottom panel of Fig.\,\ref{fig:whiidep}. The effect of the higher
radiative transition rates for \whii{} = 0.5 is evidenced in the larger level populations
for the higher energy states compared to the case of \whii{} = 0.05.

\subsection{Beaming}
\label{section:beaming}
\begin{figure}
  \begin{center}
    \includegraphics[width=\columnwidth]{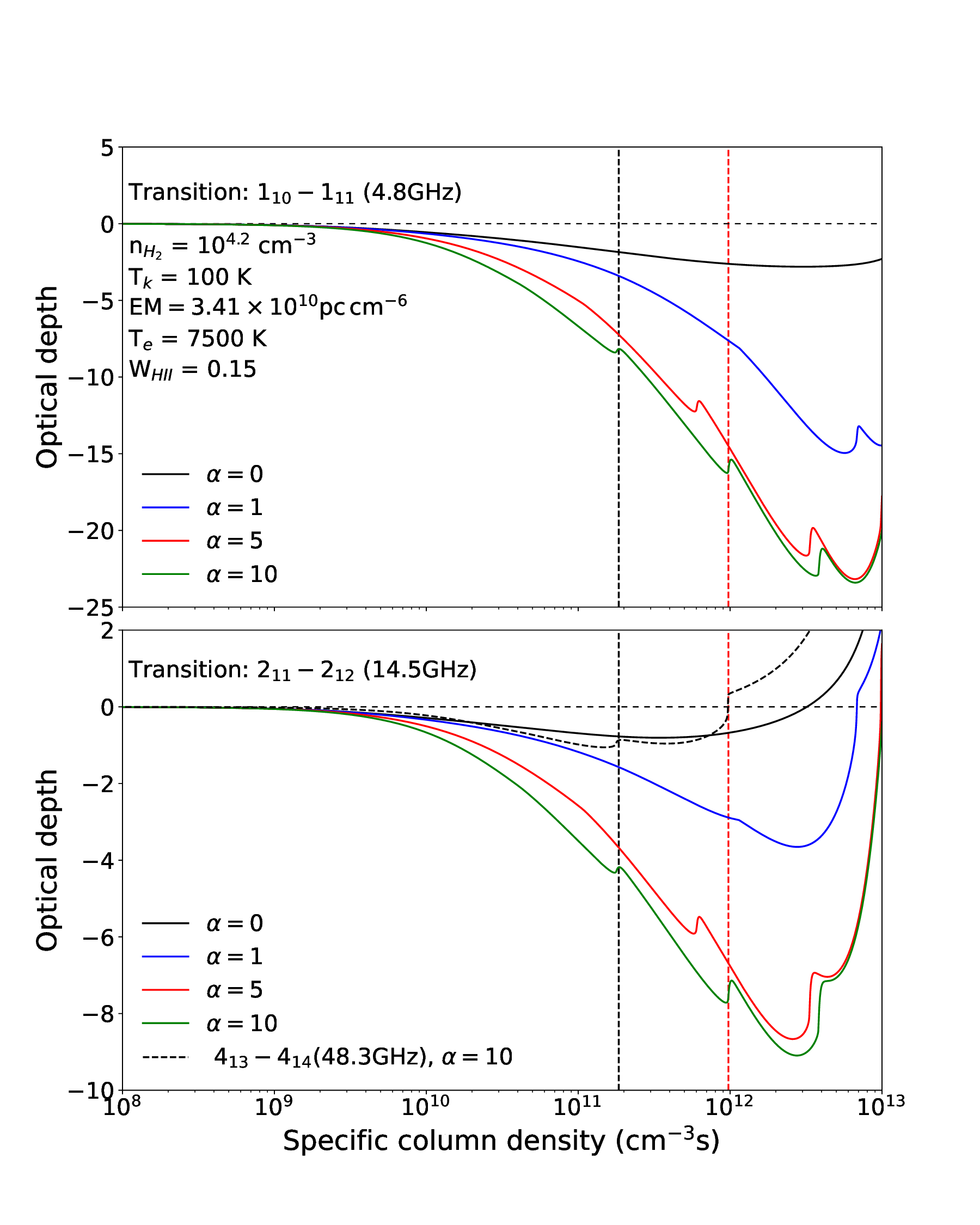}
  \caption{Illustration of the effect of beaming on the optical depth of the \gstate{} and
  \fstate{} transitions. The parameters for the calculation are shown in the upper
  panel. The red dashed vertical line indicates the specific column density where the
  \fourstate{} transition (48 GHz) switches from being inverted to non-inverted. The black
  dashed vertical line indicates the specific column density where the maser pathlength
  equals $7 \times 10^{16}$ cm.}
    \label{fig:beaming01}
    \end{center}
\end{figure}

Inspection of Figs.\,\ref{fig:tautex01}, \ref{fig:emdep01} and \ref{fig:emdep02} shows
that the maximum negative optical depth for the \gstate{} transition is less than $-3$,
which is too small to account for observed maser brightness temperatures of the order of
$10^6$ to $10^8$ K, even if the masing region is projected against a background source
with a brightness temperature of $10^4$ K. Fig.\,\ref{fig:beaming01} compares the
variation of \taug{} (upper panel) and \tauf{} (lower panel) with specific column density
when beaming is implemented, according to Eq.\,\ref{eq:finbeaming}, for $\alpha$ = 0 (no
beaming), 1, 5, and 10. The first point to be noted is that, for $\alpha = 5$ and
$\alpha = 10$, \taug{} can be as large as $\sim -23$ and \tauf{} $\sim -9$. However, these
values occur at specific column densities which correspond to maser pathlengths $\sim 8
\times 10^{17}$ cm, which is unrealistically large. Following \citet{Boland1981} and
\citet{Hoffman2003}, the maser pathlength is assumed to be geometrically constrained by
the radius of the \hii{} region. Using the data of \citet{Meng2022}, an average radius,
$\langle R_{\hii{}}\rangle$, of $4.8 \times 10^{16}$ cm and with a standard deviation,
$\sigma$, of $3.0 \times 10^{16}$ cm, is found for 39 ultra-compact \hii{} regions in Sgr
B2. For the present calculations a maser pathlength of $7 \times 10^{16}$ cm was used
which is slightly smaller than $\langle R_{\hii{}}\rangle + \sigma$. In
Fig.\,\ref{fig:beaming01} such a maser pathlength corresponds to a specific column density
of $1.85 \times 10^{11}\,\mathrm{cm^{-3}\,s}$ and is indicated by the black dashed
vertical line. At this specific column density \taug{} has the values $-1.86, -3.81,
-7.56, -9.10$ respectively for $\alpha = 0,~1,~5,~10$. The corresponding values for
\tauf{} are $-0.77,-1.79,-3.85,-4.69$. These numbers clearly illustrate the effect of
beaming.

\begin{figure}
  \begin{center}
    \includegraphics[width=\columnwidth]{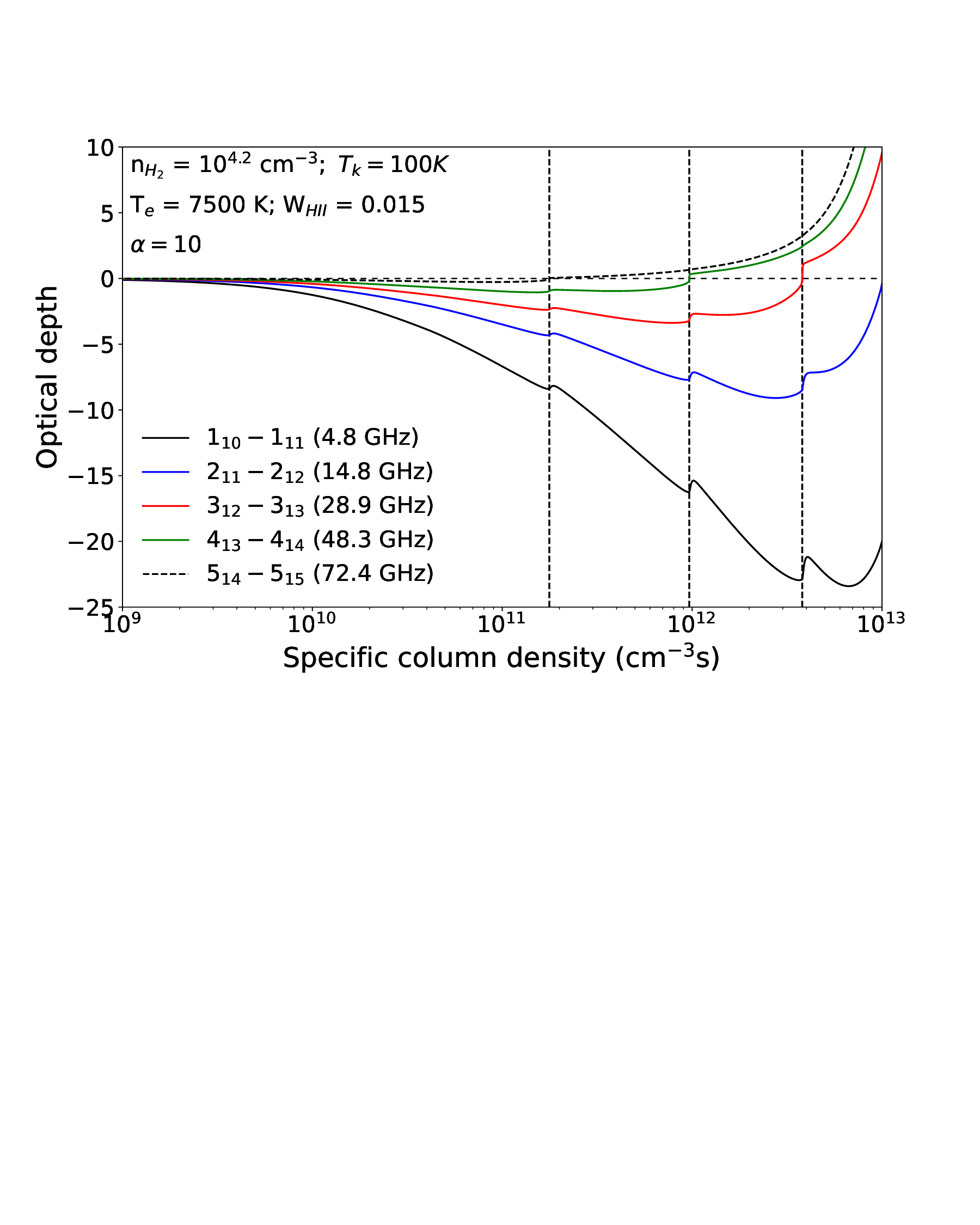}
    \caption{Illustration of the effect of beaming on variation of the optical depth of
      the first five doublet states. The parameters for the calculation are shown in the
      graph. From left to right the dashed vertical lines indicate the specific column
      densities where, respectively, the \fourstate{}, \threestate{}, and \fstate{}
      transitions switch from being inverted to not inverted.}
    \label{fig:beaming01a}
    \end{center}
\end{figure}

\begin{figure}
  \begin{center}
    \includegraphics[width=\columnwidth]{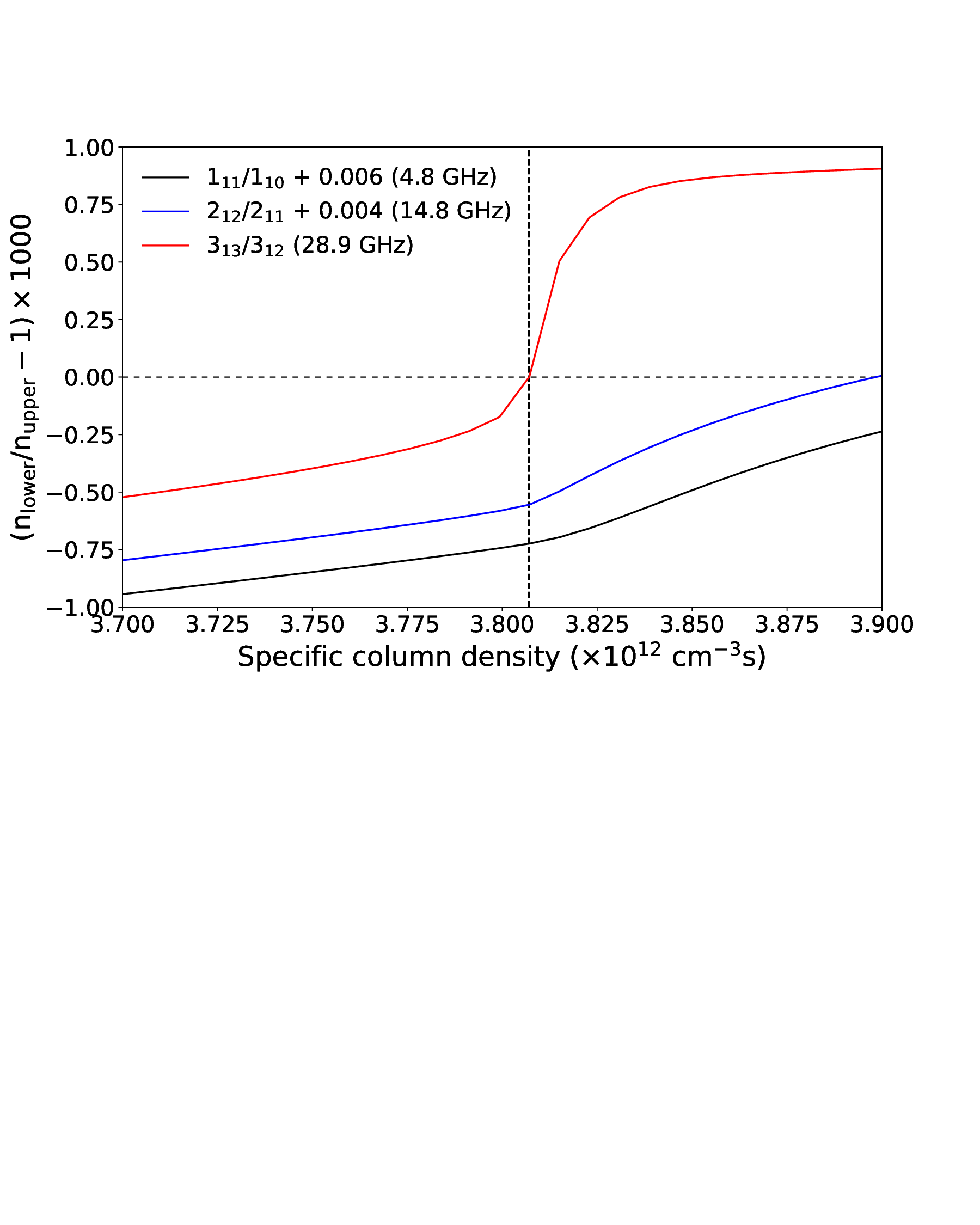}
    \caption{ Illustration of the change in the ratio, $\mathrm{n_{lower}/n_{upper}}$, of
      populations of the lower to upper levels for the \gstate{}, \fstate{} and,
      \threestate{} transitions around the specific column density where the \threestate{}
      transition switches from inversion to non-inversion at the position indicated by the
      dashed vertical line.}
    \label{fig:beaminglevpop}
    \end{center}
\end{figure}

A striking feature in the behaviour of the optical depths in Fig.\,\ref{fig:beaming01} is
the small but rapid changes in the optical depth with increasing specific column density
for $\alpha = 5$ and $\alpha = 10$. For example, it is seen that, for $\alpha = 10$, both
the \gstate{} and \fstate{} transitions show a small decrease (becoming less negative) in
optical depth at a specific column density of $10^{12}\,\mathrm{cm^{-3}\,s}$, after which
it further increases (becoming more negative) again. While this behaviour might at first
appears odd when compared with the smooth variation of the optical depth as shown in
Figs.\,\ref{fig:emdep01} and \ref{fig:emdep02}, in fact, it reflects the coupling between
the doublet states of o-\formaldehyde{}. This behaviour is further illustrated in
Fig.\,\ref{fig:beaming01a} where the variation of the optical depth as a function of
specific column density is shown for the first five doublet transitions as indicated in
the graph. All five transitions are inverted at very small specific column densities. The
\fivestate{} transition (dashed black line) is the first to switch from inversion to
non-inversion, followed by \fourstate{} (green) and \threestate{} (red) at larger specific
column densities. It is seen that when a higher $J$ doublet switches from being inverted
to non-inverted it has the effect to create a small ``bump'' in the optical depths of the
inverted lower $J$ doublets. To understand the origin of this effect, we show in
Fig.\,\ref{fig:beaminglevpop} the behaviour of $\mathrm{n_{lower}/n_{upper}}$ for the
\gstate{}, \fstate{} and \threestate{} transitions for specific column densities between
$3.7 - 3.9 \times 10^{12}\,\mathrm{cm^{-3}\,s}$. A transition is inverted when
$\mathrm{n_{lower}/n_{upper} < g_{lower}/g_{upper}}$, while keeping in mind that
$\mathrm{g_{upper} = g_{lower}}$ for the doublet transitions.  The \threestate{}
transition switches from inversion to non-inversion at a specific column density of $3.807
\times 10^{12}\,\mathrm{cm^{-3}\,s}$, indicated by the vertical dashed black line. It is
seen that there is a rapid increase in $\mathrm{n_{3_{13}}/n_{3_{12}}}$ after switching
from inversion to non-inversion which is due to $\mathrm{n_{3_{13}}}$ increasing faster
than $\mathrm{n_{3_{12}}}$ as the specific column density increases.  Since the
\fourstate{} transition is not inverted anymore at this specific column density, it means
that $\mathrm{n_{4_{14}} > n_{4_{13}}}$. Furthermore, since the spontaneous decay rates
between the lower levels of the doublet states are greater than between the corresponding
upper levels, it follows that $3_{13}$ is populated faster than $3_{12}$ which explains
the rapid increase in $\mathrm{n_{3_{13}}/n_{3_{12}}}$ after inversion. Similarly, the
spontaneous decay rate for $3_{13} \rightarrow 2_{12}$ is greater than that for $3_{12}
\rightarrow 2_{11}$, which means that $2_{12}$ is populated slightly faster than
$2_{11}$. This gives rise to a slight decrease in the inversion of the \fstate{}
transition and therefore to the decrease of \tauf{} associated with the switching from
inversion to non-inversion of the \threestate{} transition. The behaviour of \taug{}
follows directly from similar reasons as for \tauf{}.  The important point of this
behaviour is that it illustrates the coupling between the doublet states and how changes
from inversion to non-inversion, and vice versa, of a higher $J$ doublet state affects
doublet states at lower $J$. The reason why this behaviour is present only for stronger
beaming can easily be argued from the effect that beaming has on the escape probability
for inverted transitions and therefore on the level populations which, in turn, determines
the source function and therefore the internal radiation field.

\subsection{Variation of $\tau_{4.8}$ and $\tau_{14.5}$ on the $\mathrm{n_{H_2}-T_k{}}$ plane}

\begin{figure}
  \begin{center}
    \includegraphics[width=\columnwidth]{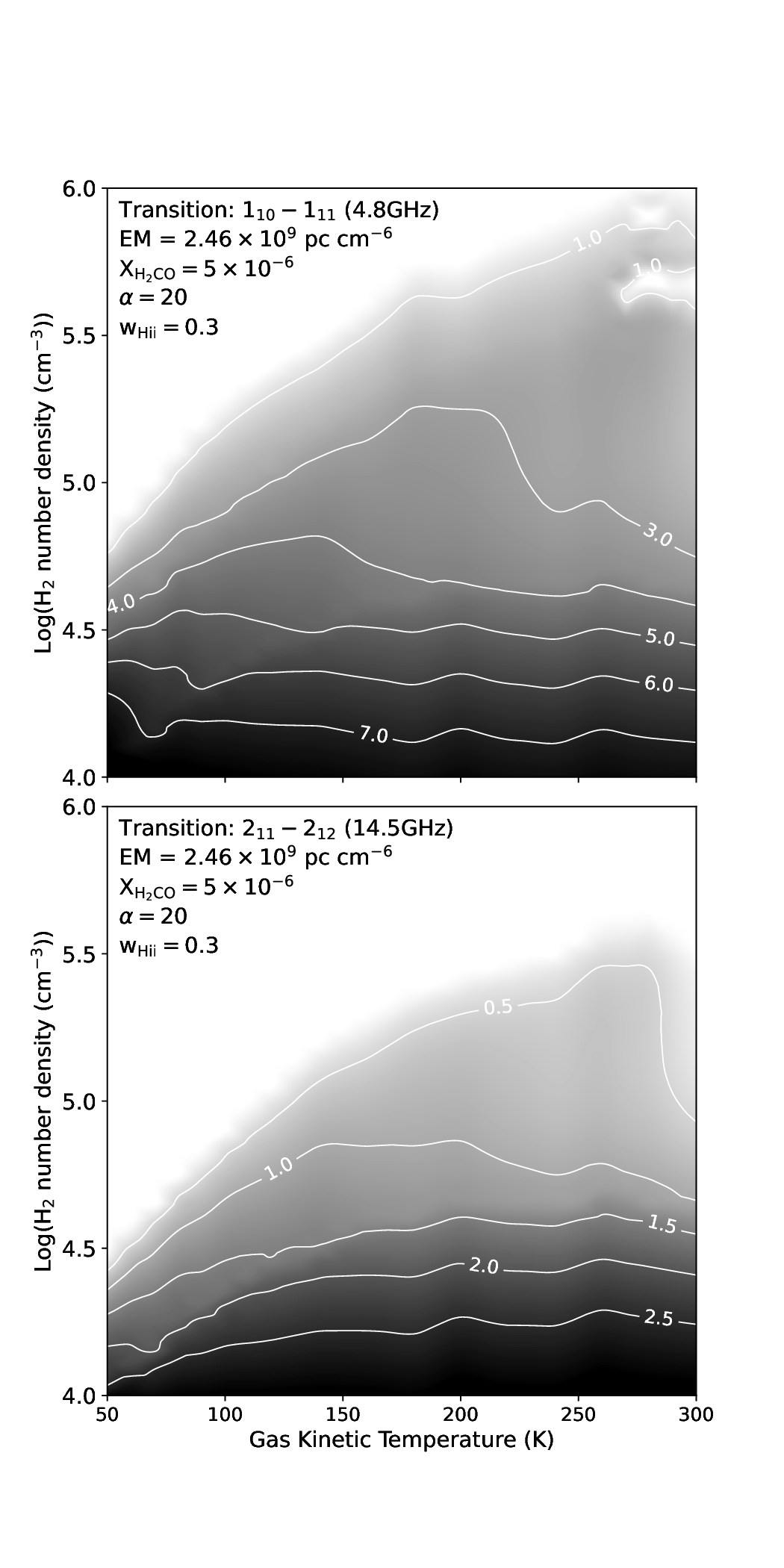}
    \caption{ Example of the variation of $\lvert\tau_{4.8}\rvert$ (upper panel) and
      $\lvert\tau_{14.5}\rvert$ (bottom panel) on the \nhtwo{}-\tk{} plane for EM = $2.46
      \times 10^9\,\mathrm{pc\,cm^{-6}}$ and $\alpha = 20$. The case for $\alpha = 10$
      does not differ substantially from what is shown.  }
    \label{fig:nh2tk01}
    \end{center}
\end{figure}

\begin{figure}
  \begin{center}
    \includegraphics[width=\columnwidth]{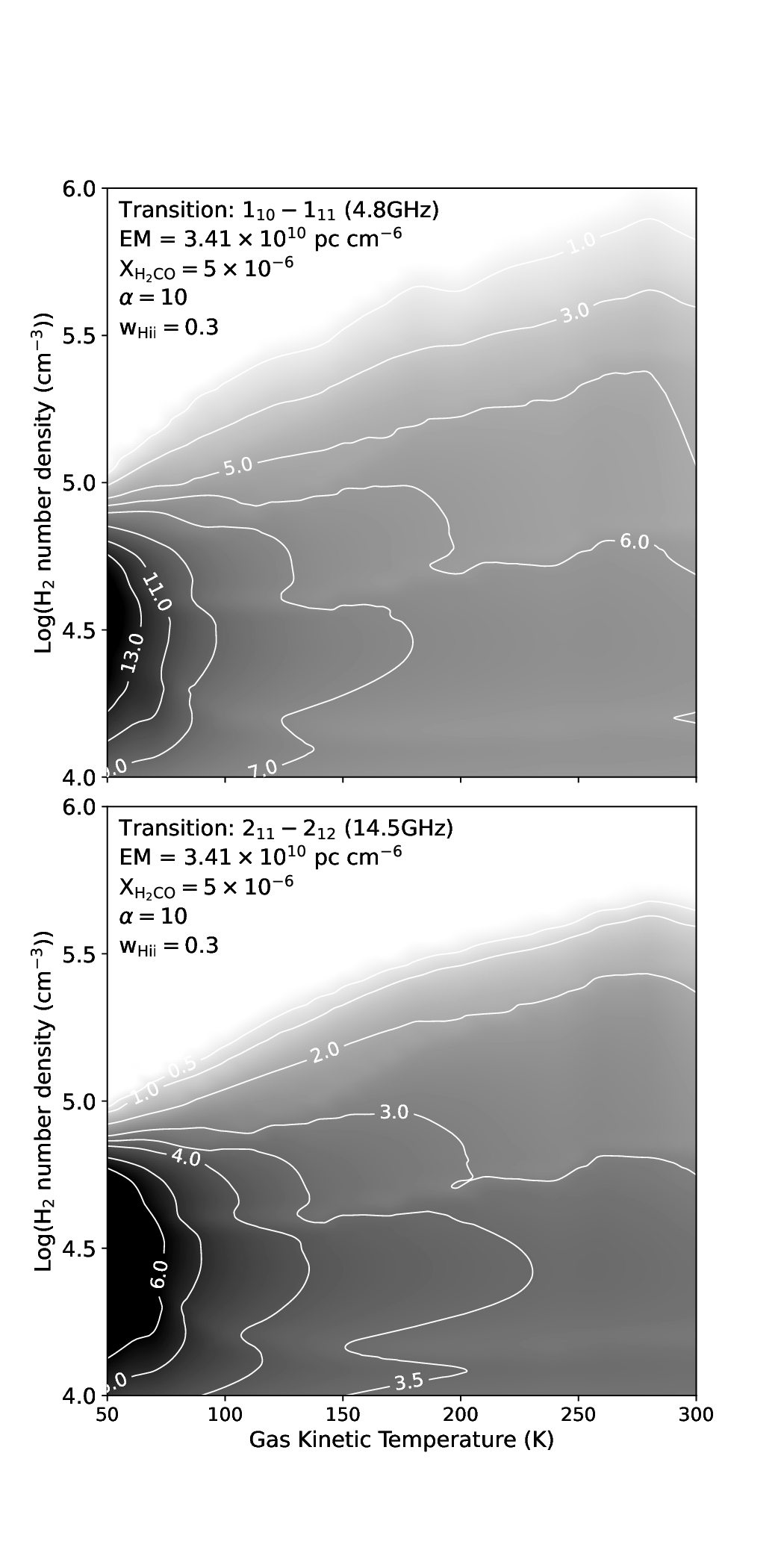}
    \caption{Example of the variation of $\lvert\tau_{4.8}\rvert$ (upper panel) and
      $\lvert\tau_{14.5}\rvert$ (bottom panel) on the \nhtwo{}-\tk{} plane for EM = $3.41
      \times 10^{10}\,\mathrm{pc\,cm^{-6}}$ and $\alpha = 10$.
    }
    \label{fig:nh2tk02}
    \end{center}
\end{figure}

\begin{figure}
  \begin{center}
    \includegraphics[width=\columnwidth]{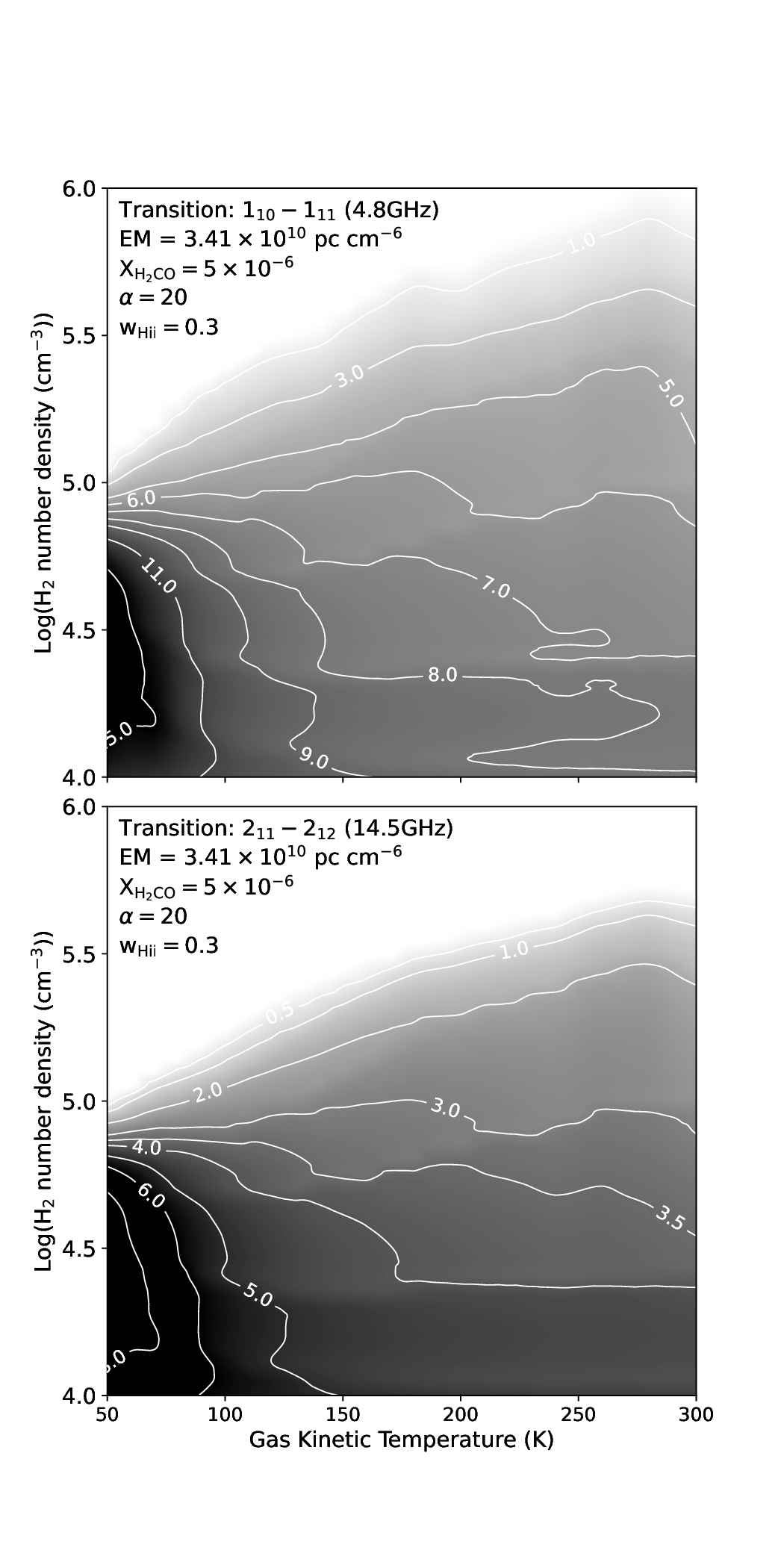}
    \caption{Same as for Fig.\,\ref{fig:nh2tk02} but for $\alpha = 20$.
    }
    \label{fig:nh2tk03}
    \end{center}
\end{figure}

\begin{figure}
  \begin{center}
    \includegraphics[width=\columnwidth]{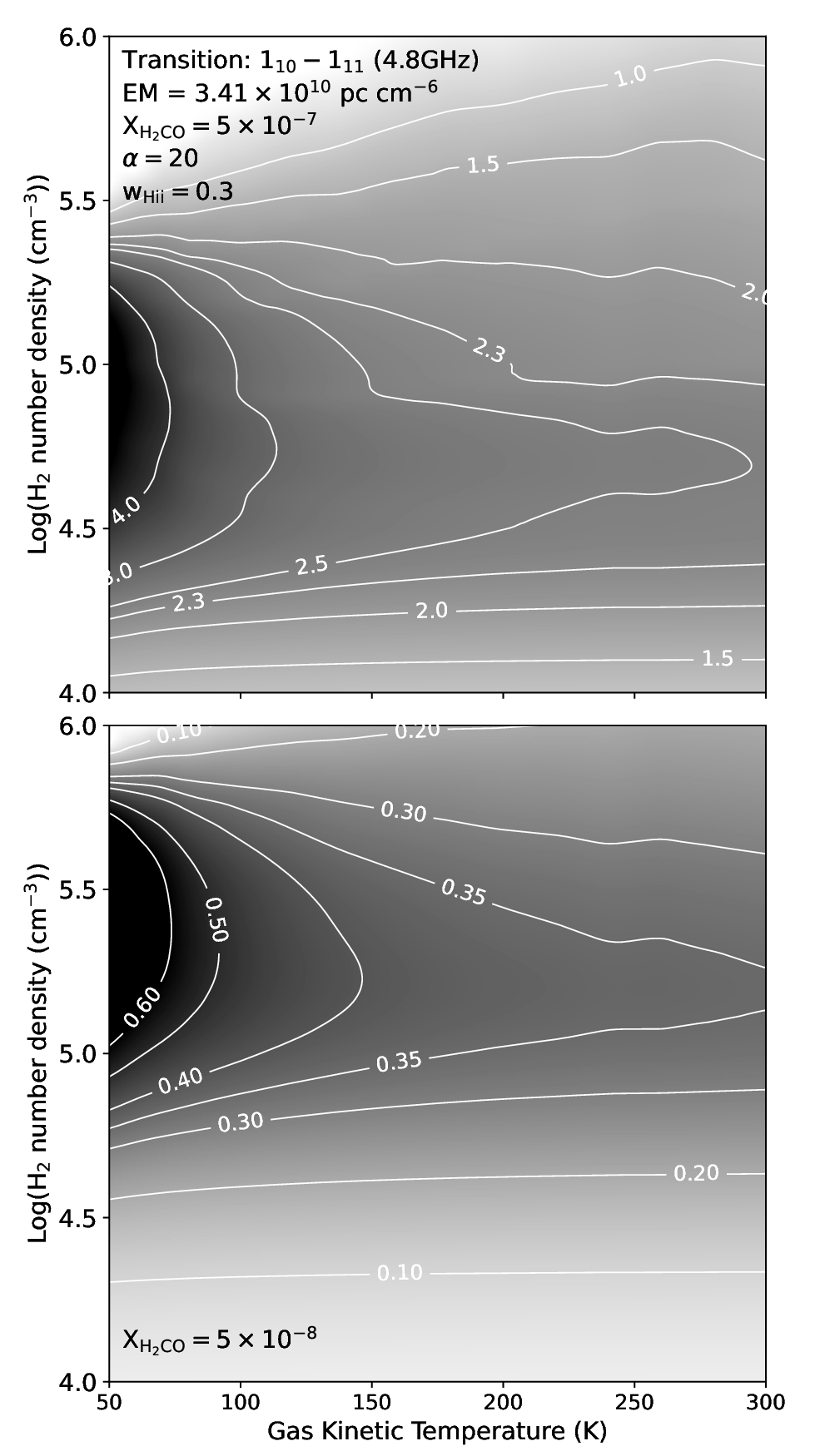}
    \caption{Same as in the upper panel for Fig.\,\ref{fig:nh2tk02} but for (a)
      $\mathrm{X_{H2CO}} = 5 \times 10^{-7}$ in the upper panel and (b) $\mathrm{X_{H2CO}} = 5
      \times 10^{-8}$ in the lower panel.  }
    \label{fig:nh2tk04}
    \end{center}
\end{figure}

Figures \,\ref{fig:nh2tk01}-\ref{fig:nh2tk04} show a small selection of examples of the
variation of \taug{} and \tauf{} on the \nhtwo-\tk{} plane that illustrate the typical
variation of these two quantities. The maser pathlength used for these results was $7
\times 10^{16}$ cm. It is seen that, in the presence of a free-free radiation field,
inversion of the \gstate{} and \fstate{} transitions can occur over almost the whole of
the \nhtwo{}-\tk{} plane for $50 \le T_k \le 300$ K and $10^4 \le n_{H_2} \le
10^6~\mathrm{cm^{-3}}$. The regions in white on the \nhtwo{}-\tk{} plane should not be
interpreted as implying that there is no inversion; in these regions inversion occurs at
maser pathlengths less than $7 \times 10^{16}$ cm with \taug{} and \tauf{} being less
than 1. It is interesting to note that for EM = $2.46 \times 10^9\,\mathrm{pc\,cm^{-6}}$
the contours for \taug{} $>$ 5 (upper panel of Fig.\,\ref{fig:nh2tk01}) are almost
independent of the gas kinetic temperature and that the largest values of \taug{} occur at
lower \htwo{} densities. The pattern changes significantly when EM = $3.41 \times
10^{10}\,\mathrm{pc\,cm^{-6}}$ (Figs.\,\ref{fig:nh2tk02} and \ref{fig:nh2tk03}). The
largest optical depths for \taug{} now are from -9 to -15 and are found at \tk{}
$\lesssim$ 100 K and \nhtwo $\lesssim 6 \times 10^4\,\mathrm{cm^{-3}}$.  As a further
comparison with the upper panel of Fig.\,\ref{fig:nh2tk01}, the \taug{} = 5 contour, for
example, has shifted to higher \htwo{} densities. An important aspect to note in the
bottom panels of Figs.\,\ref{fig:nh2tk02} and \ref{fig:nh2tk03} is that \tauf{} $> -2$
over most of the region where the \fstate{} transition is inverted and even have values $>
-6$.

The behaviour of \taug{} and \tauf{} as shown in Figs.\,
\ref{fig:nh2tk01}-\ref{fig:nh2tk03} is typical for other values of the emission measure,
beaming parameter, dilution factor and $\mathrm{X_{H_2CO}} = 5 \times 10^{-6}$.  Recent
estimates of the \formaldehyde{} abundance relative to \htwo{} for 36 hot molecular cores
\citep{Taniguchi2023}, points to abundances ranging between $\log
\mathrm{X_{H_2CO}}=-8.68$ and $\log \mathrm{X_{H_2CO}}=-6.02$ with $\overline{\log
  \mathrm{X_{H_2CO}}} = -7.23$, which translates to a value of $ 5.8 \times 10^{-8}$. In
Fig.\,\ref{fig:nh2tk04} we show the variation of \taug{} on the \nhtwo{}-\tk{} plane for
\formaldehyde{} abundances of $5 \times 10^{-7}$ and $5 \times 10^{-8}$ to illustrate the
effect of having one and two orders of magnitude smaller abundances. Even with a beaming
parameter of $\alpha = 20$ and a dilution factor of 0.3, is it seen that \taug{} is
significantly smaller than when \xform{} = $5 \times 10^{-6}$.

\section{Comments on critique against free-free radiative pumping}

The results presented above are in principle the same as that of \citet{Boland1981},
i.e. that the \gstate{} transition can be inverted by a free-free radiation
field. Although the results presented above are a significant extension beyond the work of
\citet{Boland1981}, the same criticism raised against the model of \citet{Boland1981}, may
therefore be raised against the results presented here. We, therefore, comment here on a
number of points of critique which have been raised against the free-free radiative
pumping of the \formaldehyde{} masers.

\begin{enumerate}

\item \textit{The emission measures of the associated \hii{} regions are too small:} The
  validity of the model of \citet{Boland1981} has been questioned by
  e.g. \citet{Pratap1994}, \citet{Mehringer1994} and \citet{Araya2005} on the basis of the
  emission measures of the associated \hii{} regions being too small. The fact that a
  significant number of \formaldehyde{} masers are found in the vicinity of \hii{}
  regions, and that the results of the numerical calculations presented above suggest that
  these masers may be pumped radiatively by a free-free radiation field, raises the
  question of whether ``the problem'' is with the pumping scheme or with estimating the
  emission measures of the \hii{} regions associated with 4.8 GHz \formaldehyde{} masers.
  To illustrate that the observationally derived emission measure of an \hii{} region can
  differ significantly from the real emission measure, the method followed by
  \citet{Meng2022} to determine the emission measures of \hii{} regions in Sgr B2, was
  applied to a synthetic \hii{} region calculated with the photo-ionization code Cloudy
  \citep{Ferland2017}. The following parameters were used in the Cloudy calculation:
  Blackbody temperature = 35900 K, total luminosity = $10^{38.645}\,\mathrm{erg\,s^{-1}}$,
  inner radius = $10^{12.845}$ cm, outer radius = $10^{16.7}$ cm. The radial dependence of
  the \ion{H}{I} density followed the power law $$ n(r) = n_o(r_o)\left(1 + \frac{\Delta
    r}{R_s} \right)^\alpha ~~~ \mathrm{cm^{-3}}$$ where $\Delta r$ is the distance from
  the inner radius, and $R_s$ the scale depth. The \ion{H}{I} density at the inner radius
  was $10^{6.5}~\mathrm{cm^{-3}}$, $R_s = 10^{15.3}~\mathrm{cm}$, and $\alpha = -0.7$. The
  emission measure was calculated from the resulting radial electron density distribution
  and was found to be $1.2 \times 10^{10}~\mathrm{pc\,cm^{-6}}$.

To estimate the emission measures of the \hii{} regions in Sgr B2, which contains roughly
half of the known \formaldehyde{} masers in the Galaxy, \citet{Meng2022} used the observed
6 GHz and 22.4 GHz flux densities. Following \citet{Meng2022}, the observed flux density
is related to the actual radius and emission measure of the \hii{} region according to
\begin{equation}
\frac{S_\nu}{\mathrm{mJy}} =
8.183\times10^{-4}\left(\frac{r_{\mathrm{calc}}}{\mathrm{arcsec}}\right)^2
\left(\frac{\nu}{\mathrm{GHz}}\right)^2\left(\frac{T_e}{\mathrm{K}}\right)(1 -
e^{-\tau_{\nu}})
\end{equation}
where the optical depth $\tau_{\nu}$ is
\begin{equation}
  \tau_{\nu} = 8.235 \times 10^{-2}
  \left(\frac{\nu}{\mathrm{GHz}}\right)^{-2.1}\left(\frac{T_e}{\mathrm{K}}\right)^{-1.35}
  \left(\frac{\mathrm{EM}}{\mathrm{pc\,cm^{-6}}}\right)
\end{equation}
\citep[see equations 1 and 2 of][]{Meng2022}. With the observed flux densities at 6 GHz
and 22.4 GHz, i.e. $S_6$ and $S_{22.4}$, there are two equations with two unknowns which
have to be solved for $r_{\mathrm{calc}}$ and EM. The solution for EM can be obtained by
taking the ratio $S_{22.4}/S_6$, which, after inserting the numerical values, gives
\begin{equation}
\frac{S_{22.4}}{S_6} = 13.938\frac{1 - \exp(-k_1 \mathrm{EM})}{1 - \exp(-k_2 \mathrm{EM})}
\label{eq:estem}
\end{equation}
with $k_1 = 4.7878 \times 10^{-10}$ and $k_2 = 7.6128 \times 10^{-9}$ with units of
$\mathrm{pc^{-1}\,cm^{6}}$.

\begin{figure}
\begin{center}  
  \includegraphics[scale=0.3]{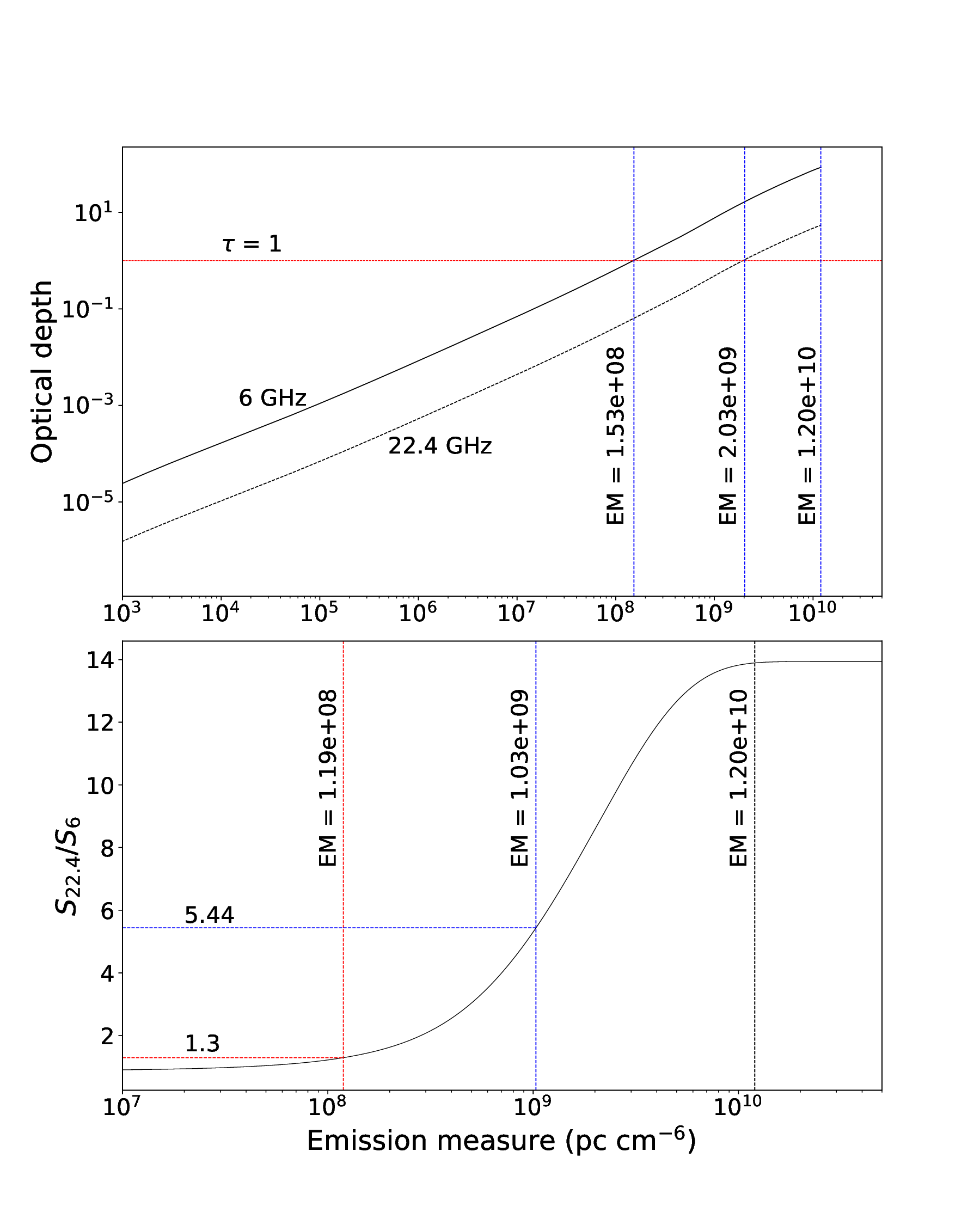}
  \end{center}
    \caption{\textit{Upper panel:} Optical depths at 6 GHz (black solid line) and 22.4 GHz
      (black dashed line) as a function of emission measure where both quantities are
      calculated inwards from the outer boudary. \textit{Lower panel:} Variation of the
      ration $S_{22.4}/S_6$ as a function the emission measure.}
    \label{fig:emgraph}
\end{figure} 

In the upper panel of Fig.\,\ref{fig:emgraph}, we show the variation of the optical depth
at 6 GHz and 22.4 GHz, calculated from the outer boundary of the synthetic \hii{} region
inwards as a function of the emission measure also calculated from the outer boundary
inwards. For 6 GHz, the optical depth reaches a value of 1 at an emission measure of
$1.53\times10^8~\mathrm{pc\,cm^{-6}}$ and for 22.4 GHz at an emission measure of $2.03
\times 10^9~\mathrm{pc\,cm^{-6}}$. The fact that the emission at 6 GHz and 22.4 GHz
originate from regions with different emission measures (both less than the true emission
measure) is already an indication that an estimate of the true emission measure must
result in an emission measure which is smaller than the true emission measure.

Estimating the emission measure using the method of \citet{Meng2022} is shown graphically
in the lower panel of Fig.\,\ref{fig:emgraph}. The S-shaped solid line represents the
ratio $S_{22.4}/S_6$ as a function of the emission measure according to
Eq.\,\ref{eq:estem}. From the emitted spectrum of the synthetic \hii{} region, it is found
that $S_{22.4}/S_6 = 5.44$, which corresponds to an emission measure $\approx 1.0 \times
10^9~\mathrm{pc\,cm^{-6}}$ as indicated on the figure. This is slightly more than an order
of magnitude smaller than the true emission measure of the synthetic \hii{} region. The
red dashed lines in Fig.\,\ref{fig:emgraph} are for source number 5 in Tables B1 and B2 of
\citet{Meng2022}, which is used as verification. For this source $S_{22.4}/S_6 = 1.3$ and
the emission measure quoted by \citet{Meng2022} is $1.19 \times
10^{8}~\mathrm{pc\,cm^{-6}}$. It is seen that the solution falls on the line representing
Eq.\,\ref{eq:estem}.

The implication of the above example for evaluating the current results, but also that of
\citet{Boland1981}, is that some of the \hii{} regions associated with \formaldehyde{}
masers may have emission measures significantly larger than what has been
estimated. Rejection of the free-free pumping of the masers on the basis that the emission
measures of the associated \hii{} regions are too small has, therefore, itself to be
assessed more critically. Inspection of the emission measures derived by \citet{Meng2022}
shows that there are five \hii{} regions with emission measures greater than
$10^9~\mathrm{pc\,cm^{-6}}$. If, for all five of these cases, the emission measures were
underestimated by one order of magnitude, it would mean that their sample contains five
\hii{} regions with emission measures greater than $10^{10}~\mathrm{pc\,cm^{-6}}$ which
would significantly change the distribution of emission measures as shown in their
Fig. 4. It is interesting to note that from 3-D modelling of the thermal dust and
free-free emission in Sgr B2, \citet{Schmiedeke2016} found five \hii{} regions with
emission measures between $1.3 \times 10^{10}$ and $3.7 \times
10^{10}~\mathrm{pc\,cm^{-6}}$. The reader is referred to Section 3.5 of
\citet{Schmiedeke2016} where these authors also point out that the emission measures and
electron densities are underestimated for optically thick \hii{} regions.

\item \textit{The masers are located too far from the nearest \hii{} region or has no
  associated continuum source:} The model of \citet{Boland1981} requires that the masing
  region be located at the edge of an \hii{} region. Quite a number of detected
  \formaldehyde{} masers are not projected against an \hii{} region but are offset from an
  \hii{} region e.g. in G29.96-0.02 \citep{Pratap1994} and NGC 7538
  \citep[][]{Andreev2017}, or have no associated continuum source, for example, in Sgr B2
  \citep{Mehringer1994,Hoffman2007}.  Using the same values for \nhtwo{}, \tk{} and \te{}
  as in Fig.\,\ref{fig:beaming01}, the behaviour of the optical depth for the \gstate{},
  \fstate{}, \threestate{} and, \fourstate{} transitions as a function of radial distance
  from the centre of the \ion{H}{II} region was examined. This was done for $\alpha
  = 10$, for EM = $3.41 \times 10^{10}\,\mathrm{pc\,cm^{-6}}$ and EM = $2.46 \times
  10^{9}\,\mathrm{pc\,cm^{-6}}$, and for a maser pathlength of $7 \times 10^{16}$ cm. The
  results are shown in Fig.\,\ref{fig:varwhii01}. First, comparing the behaviour of
  \taug{} for the two emission measures, it is seen that for EM = $3.41 \times
  10^{10}\,\mathrm{pc\,cm^{-6}}$, inversion occurs over the interval $1 \le
  r/R_{\ion{H}{II}} \lesssim 6$ with \taug{} having its maximum value at
  $r/R_{\ion{H}{II}} \simeq 1.3$. For EM = $2.46 \times 10^{9}\,\mathrm{pc\,cm^{-6}}$, the
  maximum for \taug{} is at $r/R_{\ion{H}{II}} = 1$ and inversion occurs only up to
  $r/R_{\ion{H}{II}}\sim 3$. Considering the \fstate{}, \threestate{} and \fourstate{}
  transitions, it is seen, for EM = $3.41 \times 10^{10} \,\mathrm{pc\,cm^{-6}}$, that the
  maximum $r/R_{\ion{H}{II}}$ up to which there is inversion becomes progressively smaller
  from the \fstate{} to \threestate{} to \fourstate{} transitions. For EM = $2.46 \times
  10^{9}\,\mathrm{pc\,cm^{-6}}$ it is only the \gstate{} and \fstate{} transitions that
  show significant inversion for $r/R_{\ion{H}{II}} < 3$.
\begin{figure}
  \begin{center}
    \includegraphics[width=\columnwidth]{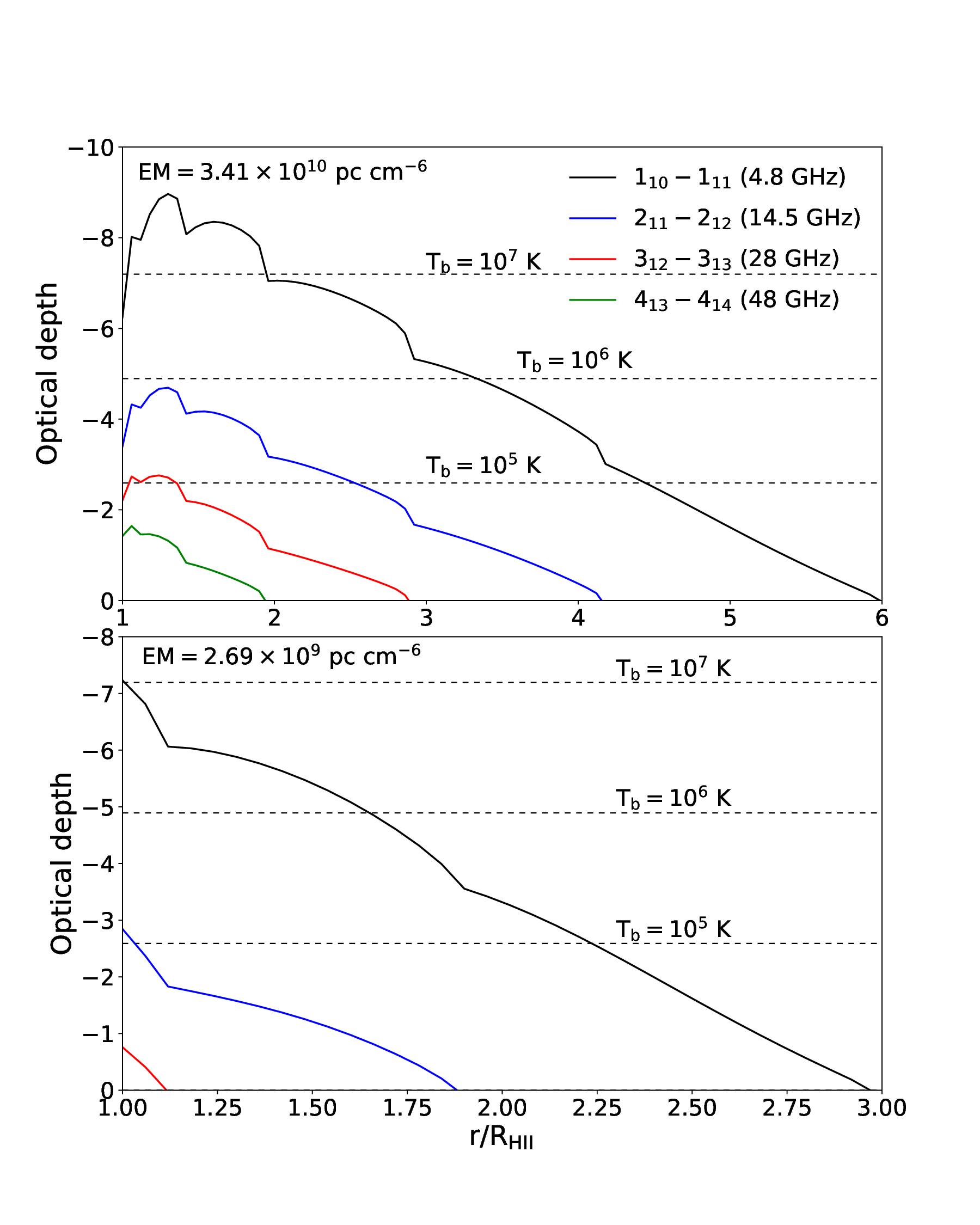}
    \caption{\textit{Upper panel:} Dependence of the optical depths of the first four
      doublet states as a function of $r/R_{\ion{H}{II}}$ for EM = $3.41 \times
      10^{10}\,\mathrm{pc\,cm^{-6}}$. The three dashed horisontal lines indicate
      theoretical brightness temperatures assuming $T_b = 7500 \exp{\lvert \tau
        \rvert}$. \textit{Lower panel:} Same as for the upper panel but for EM = $2.69
      \times 10^9\,\mathrm{pc\,cm^{-6}}$.
    }
    \label{fig:varwhii01}
    \end{center}
\end{figure}

  The results presented in Fig.\,\ref{fig:varwhii01} show that inversion of the \gstate{}
  transition can occur for distances from the centre of the \hii{} region of up to six
  times the radius of the \hii{} region. That some \formaldehyde{} masers are not
  projected onto or at the edge of an \hii{} region does not seem to be a reason to
  dismiss the pumping of the masers by the free-free continuum radiation field.

  It is also not necessary to invoke the possibility of some as yet unknown mechanism for
  the pumping of the \formaldehyde{} masers in Sgr B2 which are not associated with a
  continuum source. Specific examples are those of masers B and E in Sgr B2, which appear
  to be ``located in regions devoid of continuum emission'' \citep{Mehringer1994}
  \citep[see also Fig.1 of][]{Hoffman2007}. However, using the coordinates of the two
  masers as listed by \citet{Hoffman2007}, it is found that maser B is located
  $0^{\prime\prime}.72$ from the methanol maser MMB G0.672-0.031 \citep{Caswell2010} and
  maser E is associated with the 6.7 GHz methanol maser MMB G0.645-00.042 with an angular
  separation of $0^{\prime\prime}.36$ \citep{Caswell2010}. For maser E the primary peak in
  the \formaldehyde{} maser spectrum is at 49 $\mathrm{km\,s^{-1}}$ \citep{Mehringer1994}
  and for the methanol maser it is at $49.5\,\mathrm{km\,s^{-1}}$
  \citep{Caswell2010}. Both the class II methanol and 4.8 GHz \formaldehyde{} masers are
  exclusively associated with high-mass star forming regions \citep{Breen2013, Araya2007c,
    Araya2015} which means that both these masers point to as yet undetected high mass
  star forming regions. Inspection of the 3mm maps of Sgr B2 \citep{Ginsburg2018} shows
  that both masers B and E are associated with faint 3mm continuum sources.

\item \textit{The case of IRAS18566+0408 (G37.55+0.20)}

\citet{Araya2007c} considered the applicability of the \citet{Boland1981} model, i.e.,
whether population inversion by free-free emission, can explain the \formaldehyde{} maser
in IRAS 18566+0408 (G37.55+0.20). These authors observed this high mass star forming
region with the VLA at 6, 3.6, 1.3, and 0.7 cm. The observed SED was fitted with a
combination of optically thin thermal dust emission plus optically thin free-free emission
from an ionized jet \citep[see Fig. 2 of][]{Araya2007c}. Based on the observed SED,
\citet{Araya2007c} concluded that ``excitation by radio continuum is not a feasible
mechanism with which to explain the maser in IRAS 18566+0.0408''.

It is, however, necessary to keep in mind that, due to wavelength dependent optical depth
effects, the SED of the radiation field at the location of the maser is not necessarily
the same as the observed SED. Although there seems to be no detectable \ion{H}{II} region
associated with IRAS 18566+0408, it does not mean that no highly compact \ion{H}{II}
region is present. \citet{Keto2008} concluded that ``an accreting star begins to produce
an \hii{} region once the star reaches a mass, temperature, and luminosity equivalent to
early B, about 15 - 20 $M_\odot$''. If, as remarked by \citet{Araya2007c}, IRAS 18566+0408
is powered by a single O8 type star \citep[30.8$~\mathrm{M_\odot}$,][]{Sternberg2003}, it
seems unlikely that {\it no} \hii{} region has been excited by the 30 $\mathrm{M_\odot}$
star. According to \citet{Keto2007} an \hii{} region forms if the ionizing photon rate
exceeds the hydrogen accretion rate. Consider then, for example, the simplified case of an
O8 star spherically accreting matter at a rate of $10^{-5}\,\mathrm{M_\odot\,
  yr^{-1}}$. This corresponds to a hydrogen accretion rate of $3.8 \times
10^{44}\,\mathrm{s^{-1}}$ over the surface of the star.  For an O8 star, the ionizing
photon rate is $10^{48.75} = 5.62\times10^{48}\,\mathrm{s^{-1}}$, which is a factor of
$1.5 \times 10^4$ greater than the hydrogen accretion rate.  A young hypercompact \hii{}
region with, say, EM $\gtrsim 10^{10}~\mathrm{pc\,cm^{-6}}$, $T_e = 10^4$ K, might thus
exist but its mm/sub-mm wavelength emission may be undetectable due to the strong dust
emission. The turnover frequency of the free-free emission for such an \hii{} region is
$\approx$ 55 GHz, which implies it also is highly optically thick at centimetre
wavelengths \citep[see e.g.][]{Pratap1992,Tanaka2016}. However, the continuum radiation
field of such an \hii{} region can still be the dominant radiation field within a couple
of \hii{} region radii and be responsible for the pumping of the \formaldehyde{} masers.

It is also noted that using values of the electron density, electron temperature and
physical dimension of the jet as given by \citet{Araya2007c}, the emission measure for the
jet is found to be $2.6 \times 10^6\,\mathrm{pc\,cm^{-6}}$. This is about three orders of
magnitude less than the case of EM = $2.46 \times 10^9\,\mathrm{pc\,cm^{-6}}$ which we
have considered above. The implication is that the radiative excitation rates from
\zerostate{} into the ladder of lower doublet states as well as the transition rates
within the ladder of lower doublet states is even less than that for the far-infrared dust
radiation field (Fig.\,\ref{fig:emh2co}). It can therefore be concluded that the maser is
not associated with the jet.

\item \textit{The inversion of the \gstate{} transition is due to a rare collisional
  excitation:} Based on the work of \citet{Hill1978}, \citet{Hoffman2003} proposed that
  the rarity of the \formaldehyde{} masers may be related to a small number of \hii{}
  regions for which the dissociation wave and the ionization shock wave evolve separately
  to create a shock/dissociation transition region in which rare transient species are
  produced. Within this scenario, the inversion of the \gstate{} transition is the
  achieved through the collision of \formaldehyde{} with such a rare
  species. \citet{Hoffman2003} did not state what this rare molecule or ``hot'' atom might
  be, but if it is rare, its abundance in the transition region must be significantly less
  than that of \htwo{} and \ion{H}{I} in the transition region. Furthermore, since the
  photodissociation of \formaldehyde{} can be caused by photons with energies greater than
  about 3.4 eV ($\lambda < 360\,\mathrm{nm}$) \citep{Federman1991}, the possibility of the
  photodissociation of \formaldehyde{} must therefore also be taken into account.

  Even if such a rare transient species is produced, it is reasonable to argue that for
  such a collisional process to produce an inversion of the \gstate{} transition, the
  collisional excitation rates between the levels of the lower doublet states ($\Delta J =
  1, \Delta K_c = 1$) should be comparable to or larger than the corresponding
  radiative rates to possibly be the dominant pumping mechanism. The collisional
  excitation rate (probability per unit time) is given by
  \begin{equation}
    \Gamma^c_{lu} = n_ZC_{lu} = n_Z\frac{g_u}{g_l}C_{ul}e^{-\Delta E_{ul}/kT_k}
    \label{eq:colrates}
  \end{equation}
where $Z$ indicates the rare species, $C_{ul}$ is the collisional de-excitation
coefficient and $n_Z$ the number density of the rare species. Equating the right hand
sides of Eqs.\,\ref{eq:transitionrates} and \ref{eq:colrates} it follows that the
collisional de-excitation rate required to have the collisional and radiative rates the
same is given by
\begin{equation}
  C_{ul} = \frac{1}{n_Z}\frac{c^3}{8\pi h \nu^3_{ul}}A_{ul}U_{ul}e^{\Delta E_{ul}/kT_k}
\end{equation}
Consider, for example, the case where EM = $3.41 \times 10^{10}\,\mathrm{pc\,cm^{-6}}$
with \whii{} = 0.036, as in the upper panel of Fig.\,\ref{fig:emdep02}. Even if the ``rare
species'' has a number density as high as $10^3\,\mathrm{cm^{-3}}$, it is found that
$C_{ul}$ must have values ranging from $1.25 \times 10^{-6}\,\mathrm{cm^{3}\,s^{-1}}$ to
$2.67 \times 10^{-6}\,\mathrm{cm^{3}\,s^{-1}}$ for the first eight $\Delta J = 1, \Delta
K_c = 1$ transitions in the lower ladder of doublet states. This is four orders of
magnitude larger than the corresponding collisional de-excitation rates for collisions
with \htwo{} which will indeed require a very special collision partner for
\formaldehyde{}.

\item \textit{The non-detection of the 14.5 GHz maser} As pointed out, the numerical
  results presented above suggest that under most physical conditions, the 14.5 GHz maser
  should accompany the detection of the 4.8 GHz maser. Authors such as, for example
  \citet{Hoffman2003}, explicitly searched for 14.5 GHz maser emission toward NGC 7538 and
  G29.96-0.02. According to these authors their surveys of these two star forming regions
  had a 3$\sigma$ brightness temperature sensitivity of 43 K and that a 14.5 GHz maser
  with $\tau = -1$ would produce an emission line of brightness temperature greater than
  100 K. On the basis of this non-detection, \citet{Hoffman2003} concluded that there is
  no inversion of the \fstate{} transition in these two sources. All the results presented
  above indicate that, in the presence of a free-free radiation field, the \fstate{}
  transition should be inverted, as was also found by \citet{Boland1981}.  It is difficult
  to see, for example, how to have \taug{} $\gtrsim -5$ while there is no inversion of the
  \fstate{} transition, especially when beaming plays a role. Inversion of the \fstate{}
  and higher $J$ doublet transitions can be reduced by increasing the escape probability
  for these transitions. This can be done, for example, by also setting
  $\Omega_\nu(\tau_\nu) \propto (\nu/\nu_0)^\delta$, where $\nu_0 =$ 4.8 GHz. However, due
  to the coupling of the doublet states as described above, the effect of doing this was
  found to also reduce the inversion of the \gstate{} transition.
\begin{figure}
  \begin{center}
    \includegraphics[width=\columnwidth]{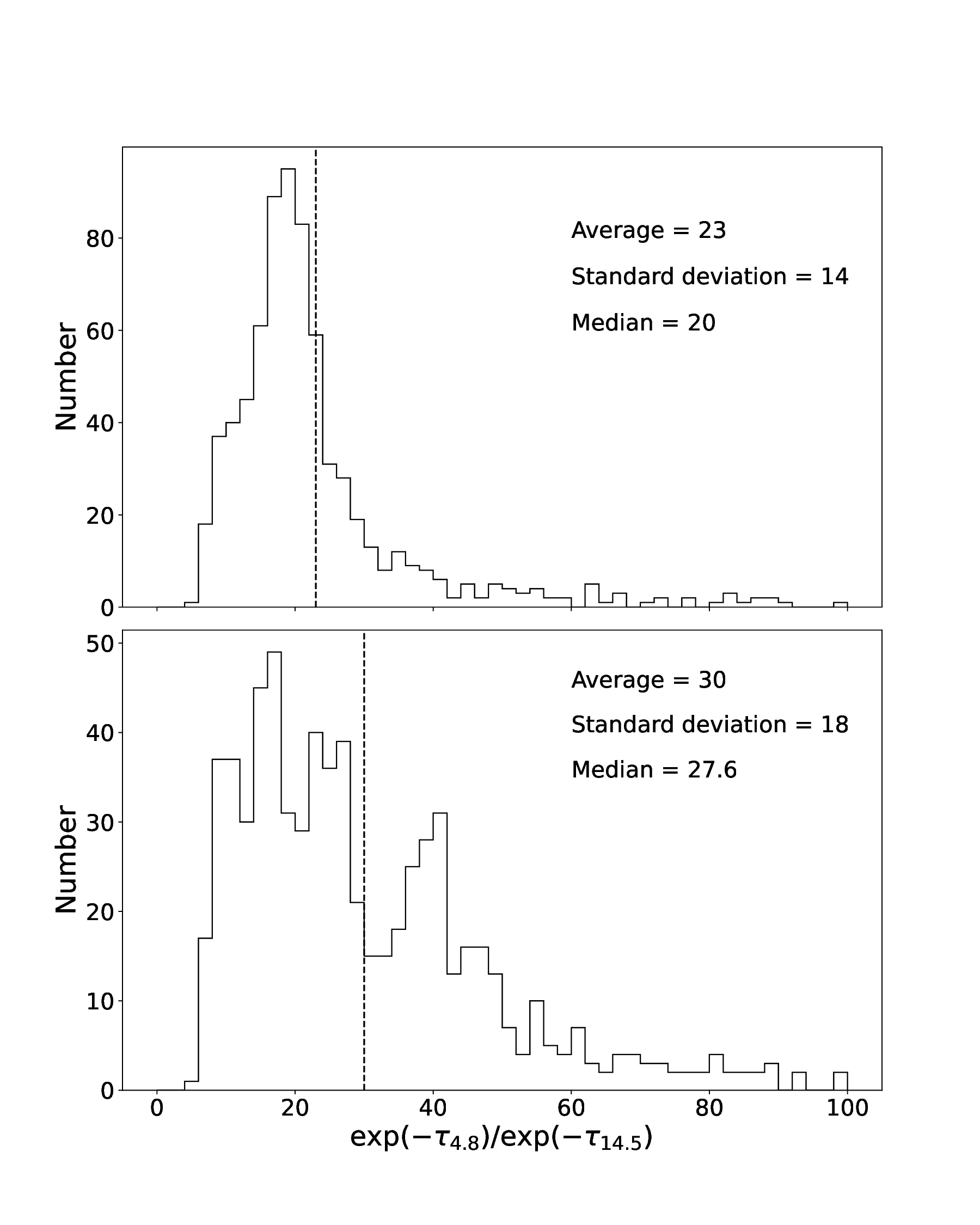}
    \caption{Histograms to show the distribution of the ratio
      $\exp(-\tau_{4.8})/\exp(-\tau_{14.5})$. The upper panel is for the case shown in
      Fig.\,\ref{fig:nh2tk02} and the lower panel for Fig.\,\ref{fig:nh2tk03}. The
      vertical dashed lines indicate the average values.}
    \label{fig:ampsratio}
    \end{center}
\end{figure}

In spite of the fact that the \fstate{} transition is almost always inverted when the
\gstate{} transition is inverted, is it worth noting that the brightness temperature of
the 14.5 GHz maser can be considerably less than that of the 4.8 GHz maser. To illustrate
this, we first show in Fig.\,\ref{fig:ampsratio} the distribution of
$\exp(-\tau_{4.8})/\exp(-\tau_{14.5})$ for the cases presented in Figs.\,\ref{fig:nh2tk02}
and \ref{fig:nh2tk03} for $\exp(-\tau_{4.8})/\exp(-\tau_{14.5}) < 100$. The average,
standard deviation and median as shown in both panels were calculated only for
$\exp(-\tau_{4.8})/\exp(-\tau_{14.5}) < 100$, since the fewer very large values ($> 1000$)
considerably skew the average and standard deviation from the main distribution. For
unsaturated gain, the distributions as shown in the upper and lower panels also reflect
the ratio of the brightnesses of the two masers for equal background intensities or
brightness temperatures. Comparison of the two distributions suggest that stronger beaming
($\alpha = 20$ (lower panel) compared to $\alpha = 10$ (upper panel)) shifts the average
23 to 30, suggesting that stronger beaming acts in favour of the 4.8 GHz maser.

However, the non-detection of the 14.5 GHz maser cannot be explained even for
$\exp(-\tau_{4.8})/\exp(-\tau_{14.5}) = 100$. Apart from the fact that, within the
framework of our calculations, $\exp(-\tau_{4.8}) > \exp(-\tau_{14.5})$, projection of the
masing region against an \ion{H}{II} region might be such that the background brightness
temperature at 4.8 GHz ($T_{4.8}$) is greater than at 14.5 GHz ($T_{14.5}$). Quite
generally the electron density is a decreasing function of the radial distance from the
centre of the \ion{H}{II} region. The effective emission measure along a chord offset from
the centre of the \ion{H}{II} region, decreases with increasing projected distance from
the centre of the \ion{H}{II} region not only due to a decrease in the electron density
but also due to the shorter chord length. The turnover frequency, $\nu_0$, also shifts to
lower frequencies since $\nu_0 \propto \mathrm{EM}^{0.476}$.  Now $T_b = T_e$ for $\nu
\leq \nu_0$ while, as a reasonable approximation, $T_b(\nu) = T_e(\nu_0/\nu)^{-2}$ when
$\nu > \nu_0$. Thus, if $\nu_0 \lesssim$ 1 GHz, the ratio between the background
brightness temperatures at 4.8 and 14.5 GHz is such that $T_{4.8}/T_{14.5} = 9.13$.
Considering therefore $T_{4.8}\exp(-\tau_{4.8})/T_{14.5}\exp(-\tau_{14.5}) =
9.13\exp(-\tau_{4.8})/\exp(-\tau_{14.5})$, it follows from Fig.\,\ref{fig:ampsratio} that
the 4.8 GHz maser may be more than two orders of magnitude brighter than the 14.5 GHz
maser.

To illustrate this with a numerical example, consider the above results of the Cloudy
calculations of a synthetic \ion{H}{II} region with EM = $1.2 \times
10^{10}\,\mathrm{pc\,cm^{-6}}$ and assume the maser to be projected against the
\ion{H}{II} region at a distance of $4 \times 10^{16}$ cm from the centre. From the Cloudy
calculations it is found that the electron density at that distance from the centre of the
\ion{H}{II} region is $\sim 5900\,\mathrm{cm^{-3}}$ and \te{} = 7165 K. For a spherical
\ion{H}{II} region the corresponding chord length is $6 \times 10^{16}$ cm with an average
electron density of $5515\,\mathrm{cm^{-3}}$ along the chord. Although the true emission
measure of the \ion{H}{II} region is $1.2 \times 10^{10}\,\mathrm{pc\,cm^{-2}}$, the
emission measure of the region against which the maser is projected is $5.9 \times
10^5\,\mathrm{pc\,cm^{-6}}$, assuming that $n_e = 5515\,\mathrm{cm^{-3}}$ and $T_e = 7165$
K is constant along the length of the chord. The corresponding turnover frequency,
$\nu_0$, for the free-free emission from such a thermal hydrogen plasma is 565 MHz.  Using
$T_b(\nu_0) = T_e$ we have $T_b(4.8\, \mathrm{GHz}) \approx 99$ K and $T_b(14.5\,
\mathrm{GHz}) \approx 10$ K. With, for example, \nhtwo{} = $10^{4.5}\,\mathrm{cm^{-3}}$,
\tk{} = 100 K, EM = $1.2 \times 10^{10}\,\mathrm{pc\,cm^{-6}}$, beaming factor $\alpha =
10$ and, \whii{} = 0.075, we find \taug{} = 4.9 and \tauf{} = 1.3 for a maser length of $7
\times 10^{16}$ cm. This gives a 4.8 GHz maser brightness temperature of 13335 K and a
14.5 GHz maser brightness temperature of only 40 K. Although this example cannot be
considered as presenting a full explanation of the (apparent) absence of the 14.5 GHz
masers, it does illustrate that the brightness temperature of the 14.5 GHz maser can be
significantly less than that of the 4.8 GHz maser and that the projection of the masing
region against the background \ion{H}{II} may be a contributing factor.

\end{enumerate}

\section{Evaluation and conclusions}

As stated in the Introduction, the primary aim of this work is to first establish
whether the pumping scheme proposed by \citet{vanderwalt2022} is indeed also responsible
for the inversion of the \gstate{} transition by a free-free radiation field. The results
presented in Section\,\ref{section:pumping} convincingly showed the transfer of molecules
from the ladder of lower doublet states to the ladder of upper doublet states is essential
to create an inversion of the \gstate{} transition. Since inversion of the \gstate{}
transition can also be achieved through collisions alone through the same process
\citep{vanderwalt2022}, it follows that any other proposed pumping mechanism must operate
in the same way, i.e., a faster excitation into the ladder of lower doublet states
compared to the ladder of upper doublet states, as well transfering molecules from the
ladder of lower doublet states to the ladder of upper doublet states at a rate that will
result in an inversion of the \gstate{} transition.

The results presented above clearly indicate that the typical free-free radiation field of
an \hii{} region associated with young high mass stars is very effective to invert the
\gstate{} and other doublet transitions over a wide range of \htwo{} densities and kinetic
temperatures. Depending on the emission measure of the \hii{} region, inversion can be
achieved at distances from the centre of the \hii{} region of up to six times the radius
of the \hii{} region. Our results also showed that with beaming, sufficient amplification
factors can be achieved.  Considering these results, there is, therefore, no reason to
question the viability of the pumping of the 4.8 GHz masers by a free-free radiation
field. In fact, even though the \gstate{} transition can be inverted by collisions alone,
the results clearly indicate that, closer to the \hii{} region, radiative pumping by the
free-free radiation field, by far dominates the inversion. The implication is that any
other postulated collisional process for the pumping of the masers, will have to compete
with the radiative pumping by the free-free radiation field.

As stated in the Introduction, to explain the rarity of the \formaldehyde{} masers a
distinction must be made between the pumping scheme and other possible factors that
determine the number and life-time of the masers. The fact that not only the \gstate{} but
also the \fstate{} transition can be inverted over most of the \nhtwo{}-\tk{} plane would
suggest that there should be significantly more \formaldehyde{} masers than what has been
discovered up to now. There are at least two such external factors that can possibly
contribute to the small number of detected \formaldehyde{} masers. The first is the rate
at which the \ion{H}{II} region, and therefore also its emission measure, evolves. As is
shown in Fig.\,\ref{fig:emh2co}, excitation from the \zerostate{} state into the ladder of
lower doublet states is through frequencies in the optically thin part of the spectral
energy density distribution. Since for the optically thin part the spectral energy density
is proportional to the emission measure, the excitation rates into the ladder of lower
doublet states will decrease as the emission measure decreases due to the evolution of the
\ion{H}{II} region. The evolution of the \ion{H}{II} region may therefore affect the
life-time and thus the number of \formaldehyde{} masers.

Another factor that may play a role in the number of detected \formaldehyde{} masers is
the \formaldehyde{} abundance. It is clear from Fig.\,\ref{fig:nh2tk04} that the
\formaldehyde{} abundance is relevant in determining whether the amplification
factor is large enough to produce a detectable maser. Considering again the results of
\citet{Taniguchi2023}, it is found for the sample of 36 hot cores that
$\overline{\mathrm{log X_{H_2CO}}} = -7.23$ and $\sigma = 0.62$. The skewness of the
sample is $\sim -0.4$ which indicates that it is slightly skew toward smaller abundances.
Assuming that the distribution of $\log \mathrm{X_{H2CO}}$ is Gaussian (for which the
skewness is 0), it is found that the probability is 2.4\% to find a hot core for which
$\mathrm{log X_{H_2CO}} > -6$. If the \formaldehyde{} and the class II \methanol{} masers
are associated with the same population of $\sim 1000$ high mass star forming regions
detected in the MMB survey \citep{Green2012}, it follows that there should be about 24 4.8
GHz \formaldehyde{} masers if all masers for which $\mathrm{X_{H_2CO}} > 10^{-6}$ are
detectable. This example should not be interpreted as a prediction of the total number of
4.8 GHz \formaldehyde{} masers but only to illustrate that external factors which are
independent of the pumping scheme, as as well as of the pumping mechanism (the free-free
radiation field), can lead to a relatively small number of detected maser sources. The
number may be even lower if the duration of the masing phase is also affected by the
evolution of the \ion{H}{II} region.

There are a number of 4.8 GHz \formaldehyde{} masers that also have associated 6.7 GHz
\methanol masers for which there appears to be very good spatial and velocity agreement
between the two maser species. Masers B and E in Sgr B2 already refered to above are such
examples. A striking example is in the case of G339.98-1.26 where three of the 4.8 GHz
\formaldehyde{} features correspond in velocity to three of the 6.7 GHz \methanol{} maser
features \citep{Chen2017a}. Also in the case of G0.38+0.04 is there very good spatial and
velocity agreement between the 4.8 GHz \formaldehyde{} and 6.7 GHz \methanol{} masers
\citep{Ginsburg2015}. The question then is to what extent there is similarity in the
conditions to excite the two maser species. We note the following differences which need
to be resolved if these two maser species are to be associated with the same physical and
kinematic structures. From Fig.7 of \citet{Cragg2002} it is seen that to explain the
brightness temperatures of the order of $10^{12}$ K of the 6.7 GHz \methanol{} masers
requires dilution factors, \whii{}, to be less than about $10^{-4}$. This implies $r
\gtrsim 50R_{\ion{H}{II}}$ (Eq.\,\ref{eq:dilute1}) while $r \lesssim 6R_{\ion{H}{II}}$
seems to be required for the 4.8 GHz \formaldehyde{} masers
(Fig.\,\ref{fig:varwhii01}). \citet{Cragg2002} also concluded that the brightest 6.7 GHz
\methanol{} masers are found for $10^5 < \mathrm{n_{H_2}} < 10^{8.3}\,\mathrm{cm^{-3}}$
with $\mathrm{T_k < T_d > 100\,K}$. For \formaldehyde{}, the results presented above
suggest that to have a reasonable amplification factor, e.g. \taug{} $< -5$, requires
$\mathrm{n_{H_2} < 10^{5.5}\,cm^{-3}}$ (Figs.\,\ref{fig:nh2tk01} and \ref{fig:nh2tk02}).
Although the calculations of \citet{Cragg2002} and those presented here used different
numerical methods which may contribute to some of the differences, at face value it would
seem, from a theoretical point of view, that the two maser species are not associated with
the same physical and kinematic structure in the star formation environment. 

Within the context of this discussion, it is has to be pointed out that the 4.8 GHz
\formaldehyde{} masers which are normally associated with G29.96-0.02, are most likely not
associated with the cometary \ion{H}{II} region but rather with a hot molecular core (HMC)
hosting an object identified as a high mass protostellar object \citep{Cesaroni1998,
  DeBuizer2002}. Spatially there are three water masers associated with this HMC
\citep{Hofner1996}, with one of the 4.8 GHz \formaldehyde{} masers located only
$0^{\prime\prime}.18$ from one of the water maser features. Although it is not clear what
the exact location of the 4.8 GHz \formaldehyde{} maser in the HMC is, the conclusions
drawn of it being associated with, and located toward, the edge of the cometary
\ion{H}{II} region might be different than if it is associated with a young background
HMC. The first, obviously, is the different evolutionary states of the star forming region
with which the maser is associated which also have consequences regarding the chemical
evolution. Different conclusions can also be reached regarding the optical depth required
to explain the observed brightness temperature of the maser. In the case of G29.96-0.02,
\citet{Hoffman2003} concluded that to explain the observed brightness temperature of $\sim
10^7$ K requires \taug{} $\approx -12$ based on a 6 cm background temperature of 50
K. From the upper panels of Figs.\,\ref{fig:nh2tk02} and \ref{fig:nh2tk03} it is seen
that, within the framework of the current calculations, such optical depths occur only at
lower gas kinetic temperatures. However, if the maser is projected against a hypercompact
\ion{H}{II} region associated with an earlier phase of high mass star formation, the
background temperature may be significantly higher. For example, if the background
temperature is 5000 K, the optical depth required to explain a brightness temperature of
the order of $10^7$ K, is $\sim -7.6$. It is obvious from Figs.\,\ref{fig:nh2tk02} and
\ref{fig:nh2tk03} that in this case a significantly different conclusion is reached about
the conditions under which the maser might be excited.

As a last remark we note that, having established that a free-free radiation field is
effective to pump the 4.8 GHz \formaldehyde{} masers and a far-infrared dust radiation
field is not, it follows that the underlying mechanism for the periodic variability of the
\formaldehyde{} and \methanol{} masers in IRAS18566+0408 (G37.55+0.20) cannot be due to
varying pumping conditions for both maser species. Since the pumping schemes for the two
masers are completely different, it would require an impossible degree of fine tuning of
variability between the free-free and dust radiation fields for the amplification factors
to behave in exactly the same way to produce similar light curves.  The flaring must
therefore be associated with the source of seed photons, as was argued by
\citet{vanderwalt2014} and as is strongly suggested by the recent results of
  \citet[][see their Section 6]{Gray2020}.

\section*{Acknowledgements}
I would like to thank Prof. David Modise, Dean of the Faculty of Natural and Agricultural
Sciences, as well as Prof. Francois van der Westhuizen, Deputy Dean (Research) of the
Faculty of Natural and Agricultural Sciences of the NWU, for financial support without
which this work could not have been done. I also thank an anonymous reviewer for
constructive comments to improve the paper.

\section*{Data Availability}

No new data has been generated in this project.



\bibliographystyle{mnras}
\bibliography{ref} 








\bsp	
\label{lastpage}
\end{document}